\documentclass[12pt,12pt]{article}
\usepackage{amsmath,amssymb}
\usepackage{slashed} 

\begin{document}

\def\beq{\begin{equation}}
\def\eeq{\end{equation}}
\def\beqa{\begin{eqnarray}}
\def\eeqa{\end{eqnarray}}
\newcommand{\ba}{\begin{eqnarray}}
\newcommand{\ea}{\end{eqnarray}}
\newcommand\BA{\begin{array}}
\newcommand\EA{\end{array}}

\title{\Large {\bf Exact canonically conjugate momenta approach to\\
\ \\[-12pt]
a one-dimensional neutron-proton system, I}}
\author{Seiya NISHIYAMA\footnotemark[1]~
~and
Jo\~{a}o da PROVID\^{E}NCIA\footnotemark[2]\\
\\
Centro de F\'\i sica,
Departamento de F\'\i sica,\\
\\[-10pt]
Universidade de Coimbra,
P-3004-516 Coimbra, Portugal\footnotemark[2]}

\def\bm#1{\mbox{\boldmath $#1$}}
\def\bra#1{\langle #1 |}
\def\ket#1{| #1 \rangle }

\maketitle

\vspace{1cm}

\footnotetext[1]{~$\!$Corresponding author. 

~~ E-mail address: 
seikoceu@khe.biglobe.ne.jp; nisiyama@teor.fis.uc.pt}
\footnotetext[2]{
$\!$  E-mail address:
providencia@teor.fis.uc.pt}

\vspace{0.5cm}


\begin{abstract}
$\!\!\!\!\!\!\!\!$Introducing collective variables,
a collective description of nuclear surface oscillations
has been developed with the first 
quantized language,
contrary to the second quantized one in Sunakawa's approach
for a Bose system.
It overcomes difficulties remaining 
in the traditional theories of nuclear collective motions:$\!$ 
Collective momenta are not exact 
canonically conjugate to collective coordinates and  
are not independent.    
On the contrary to such a description,
Tomonaga first gave the basic idea to approach
elementary excitations in a one-dimensional Fermi system.
The Sunakawa's approach for a Fermi system
is also expected to work well for such a problem.
In this paper,
on the $isospin$ space,
we define a density operator and further
following Tomonaga,
introduce a collective momentum.
We propose an $exact$ canonically momenta approach
to a one-dimensional neutron-proton (N-P) system
under the use of the Grassmann variables.
\end{abstract}

\vspace{0.1cm}

 
exact canonically
conjugate momenta;
Grassmann variables

\vspace{0.5cm}

PACS Number (s):
21.60.-n, 21.60.Ev


\newpage

\def\thesection{\arabic{section}}
\section{Introduction}

\vspace{-4pt}

To study quarks and $\!SU(\!N\!)\!$ colored non-Abelian gluons 
(QCD) in the large-{$\!$\it N} limit,
Jevicki and Sakita developed a collective field formalism
involving only gauge-invariant$\!$ 
variables
\cite{SJ.80a,SJ.80b,SJ.80c,SJ.80d}.
Though the QCD describes successfully short-distance 
phenomena due to the property of asymptotic freedom,
one still has the difficult problem of color confinement 
occurring as a large-distance phenomenon.$\!$
Their basic idea
consists in reformulating the quantum collective field theory 
in terms of gauge-invariant variables.
It leads to an effective Hamiltonian 
which determines the behavior in the large-{$\!$\it N} limit by classical 
stationary point solutions.

$\!\!\!\!$On the other hand, in the studies of collective motions in nuclei,
the very difficult problems of large-amplitude collective motions,
which are strongly nonlinear phenomena in quantum nuclear dynamics,
still remain unsolved.
How to go beyond the usual mean field theories towards
a construction of a theory for 
large-amplitude collective motions in nuclei
\cite{INTUW.93,Nishi.94}?

$\!\!\!\!$Applying  Tomonaga's idea for collective motion theory 
\cite{Tomo.55a,Tomo.55b}
to nuclei with the aid of the Sunakawa's discrete integral 
equation method
\cite{SYN.62},
we developed a collective description of surface oscillations of nuclei 
\cite{NishProvi.14,Nishi.77}.
It gives a possible microscopic 
foundation of nuclear collective motions related to
the Bohr-Mottelson model
\cite{BM.74}.
Introducing collective variables,
a collective description is provided by using the first 
$\!$quantized language,
$\!$contrary to the second quantized one in $\!$the Sunakawa's approach
for a Bose system.
$\!$It overcomes the difficulties still remaining 
in the traditional theoretical treatments of nuclear collective motions: 
Collective momenta in the Tomonaga's approach are not exact 
canonically conjugate to collective coordinates and  
are not independent.    
Our ${exact}$ canonically conjugate momenta to collective coordinates
are found from a viewpoint different from
the canonical transformation theory and the group theory
\cite{UiBi.70a,UiBi.70b,UiBi.70c}.
Recently
we got ${exact}$ canonical variables,
revisiting the Tomonaga's work
and described a collective motion also in two-dimensional nuclei
\cite{NishProviarXive.14}.

$\!\!\!\!$In constructing a collective field theory for 
an $SU(\!N\!)$ quantum system
\cite{BIPZ.78a,BIPZ.78b}, 
we are standing on a situation similar to the above one.
It is regarded as a common 
feature of strongly nonlinear physics.
Applying Tomonaga's idea,
a collective description of the $SU(\!N\!)$ system is plausible
in terms of collective variables invariant
under an $SU(\!N\!)$ transformation.
One of the present authors (S.N.)
gave the ${exact}$ canonical variables
using the discrete integral equation method
\cite{Nish.98},
which is regarded as a natural extension
of the Sunakawa's variables 
to the variables in the $SU(\!N\!)$ system. 
But they are derived in the first quantized language.

$\!\!\!\!$On the contrary to such collective descriptions,
to approach elementary excitations in a Fermion system,
65 years ago,
Tomonaga first gave another idea$\!$ 
\cite{Tomo.50,Emery.79}.
A $\!$similar idea $\!$was $\!$also $\!$given $\!$by $\!$Luttinger
$\!$with $\!$a slight modification of $\!$Tomonaga's proposal$\!$
\cite{Luttinger.63}.
$\!\!$Their ideas have the advantage of being exactly solvable$\!$
\cite{Solyon.79,Mahan.00}.
$\!$While the Sunakawa's approach for a Fermi system$\!$
\cite{SYN.62}
also may be anticipated $\!$to work well for such a problem.
$\!$In this paper,
on the {\it isospin} space $(T,T_{z})$,
we define a density operator $\rho^{T,T_{z}}_{k}\!$ and
further following Tomonaga,
introduce a collective momentum.
Then we propose an $exact$ canonical momenta approach
to a one-dimensional neutron-proton system
under the use of the Grassmann variables$\!$
\cite{Berezin.66,Casalbuoni.76a,Casalbuoni.76b}.

In Sec. 2
introducing collective variables $\rho^{0,0}_{k}$ 
and their associated variables $\pi^{0,0}_{k}$,
we give commutation relations among them. 
In Sec. 3,
we define {\it exact} canonically conjugate momenta 
$\Pi^{0,0}_{k}$ by a discrete integral equation
and devote ourselves to the proof
of the {\it exact} canonical commutation relation among
collective variables
$\rho^{0,0}_{k}$ and $\Pi^{0,0}_{k}$.
In Sec. 4,
the dependence of the original Hamiltonian on
$\Pi^{0,0}_{k}$ and $\rho^{0,0}_{k}$ is determined.
Section 5
is devoted to a calculation of a constant term
in the collective Hamiltonian.
Finally in Sec. 6 
some discussions and further perspectives are given.
In the Appendix
the calculation of some commutators are presented.


\newpage

\def\thesection{\arabic{section}}
\setcounter{equation}{0}
\renewcommand{\theequation}{\arabic{section}.\arabic{equation}}     
\section{Collective Variables and the Associated Relations}
Let $H$ be the Hamiltonian of a one dimensional Fermion system:\\[-12pt]
\ba 
H
\!=\!
T \!+\! V
\!=\!
\! \int \!\! dx
\psi^{\dag}(x) \!
\left( \!
-\frac{{\hbar}^{2}}{2m}
\frac{{\partial }^{2}}{\partial x^{2}} \!
\right) \!
\psi(x)
\!+\!
{\frac{1}{2}} \!
\int \!\!\! \int \!\! dx dx^{\prime}
\psi^{\dag}(x) \psi^{\dag}(x^{\prime})
V(x-x^{\prime})
\psi(x^{\prime}) \psi(x) ,
\label{Hamiltonian}  
\ea
where the field operators
$\psi(x)$ and $\psi^{\dag}(x)$
satisfy the canonical anti-commutation relations,
\ba
\{ \psi(x), \psi^{\dag}(x^{\prime}) \}
\! =\!
\delta (x \!-\! x^{\prime}) ,~
\{ \psi(x), \psi^{\dag}(x^{\prime}) \}
\! =\!
0,~
\{ \psi^{\dag}(x), \psi^{\dag}(x^{\prime}) \}
\! =\!
0 .
\label{Fermionn}  
\ea

To deal with a proton-neutron system,
let us introduce the $z$-component of $isospin$
to distinguish the Neutron and the Proton:\\[-12pt]
\ba
\left.
\BA{c}
\tau_{z}
\!=\!
\frac{1}{2} ,~~\mbox{for Neutron} , \\
\\[-6pt]
\tau_{z}
\!=\!
-\frac{1}{2} ,~~\mbox{for Proton} .
\EA
\right\}
\label{pn}
\ea
The operators ${\psi }(x)$ and ${\psi }^{\dag}(x)$ in
(\ref{Hamiltonian})
are expanded
and separated into two parts according to
(\ref{pn}),
respectively, as\\[-12pt]
\ba
\!\!\!\!
\left.
\BA{c}
{\psi }(x)
\!=\!  
{\displaystyle \frac{1}{\sqrt L}} \!
\sum_{k \tau_{z}} \!
a_{k \tau_{z}}
e^{ikx}
{\phi }_{\tau_{z}}
\!\!=\!\!   
{\displaystyle \frac{1}{\sqrt L}} \!
\sum_{k \tau_{z}} \!\!
{\displaystyle \frac{1 \!+\! 2 \tau_{z}}{2}} 
a_{k \tau_{z}}
e^{ikx}
{\phi }_{\tau_{z}}
\!\!+\!\!   
{\displaystyle \frac{1}{\sqrt L}} \!
\sum_{k \tau_{z}} \!\!
{\displaystyle \frac{1 \!-\! 2 \tau_{z}}{2}} 
a_{k \tau_{z}}
e^{ikx}
{\phi }_{\tau_{z}} \\
\\[-8pt]
{\psi }^{\dag}(x)
\!=\!  
{\displaystyle \frac{1}{\sqrt L}} \!
\sum_{k \tau_{z}} \!
a^{\dag}_{k \tau_{z}} \!
e^{-ikx}
{\phi }^{*}_{\tau_{z}}
\!\!=\!\!   
{\displaystyle \frac{1}{\sqrt L}} \!
\sum_{k \tau_{z}} \!\!
{\displaystyle \frac{1 \!\!+\!\! 2 \tau_{z}}{2}} 
a^{\dag}_{k \tau_{z}} \!
e^{-ikx}
{\phi }^{*}_{\tau_{z}}
\!\!+\!\!   
{\displaystyle \frac{1}{\sqrt L}} \!
\sum_{k \tau_{z}} \!\!
{\displaystyle \frac{1 \!\!-\!\! 2 \tau_{z}}{2}} 
a^{\dag}_{k \tau_{z}} \!
e^{-ikx}
{\phi }^{*}_{\tau_{z}} ,
\EA \!\!
\right\} 
\label{FexpaFieldop}
\ea
and the interaction potential $V(x)$
is also expanded as\\[-16pt]
\ba
\BA{c}
V(x)
\!=\!
{\displaystyle \frac{1}{L}} \!
\sum_{k }
\nu(k)
e^{ikx} .
\EA \!\! 
\label{FexpaIntPot}
\ea
Here we have used the following orthogonal relations:\\[-12pt]
\ba
\BA{c}
{\displaystyle \int} \! \left( e^{ik^{\prime}x} \right)^{*} \!\! e^{ikx} dx
\!=\!
{\displaystyle \int}  e^{i(k-k^{\prime})x} dx 
\!=\!
L \delta_{k,k^{\prime}} , ~~
{\displaystyle \int }{\phi }^{*}_{\tau^{\prime}_{z}} {\phi }_{\tau_{z}} d \tau
\!=\!
\delta_{\tau^{\prime}_{z},\tau_{z}} ,
\EA
\label{orthogonality}
\ea 
where $L$ is the length of a one-dimensional periodic box.
The anti-commutation relations among
$a_{k \tau_{z}}$'s and $a^{\dag}_{k \tau_{z}}$'s
are given as
\ba
\left.
\BA{c}
\left\{ a_{k \tau_{z}}, a^{\dag}_{k^{\prime} \tau^{\prime}_{z}} \right\}
\!=\!
\delta_{k, k^{\prime}}
\delta_{\tau^{\prime}_{z},\tau_{z}} ,\\
\\[-6pt]
\left\{ a_{k \tau_{z}}, a_{k^{\prime} \tau^{\prime}_{z}} \right\}
\!=\!
\left\{ a^{\dag}_{k \tau_{z}}, a^{\dag}_{k^{\prime} \tau^{\prime}_{z}} \right\}
\!=\!
0 .
\EA
\right\}
\label{ACR}
\ea
To study $isospin~T$ collective excitations, with the use of the Clebsch-Gordan coefficients
$
\langle
 \frac{1}{2} \tau_{z}
\frac{1}{2} \tau^{\prime}_{z} 
|T T_{z}
\rangle
$
on the $isospin$ space $(T,T_{z})$,
we define the Fourier component of the density operator
$(\rho(x) \!\!=\!\! \psi^{\dag}(x) \psi(x))$
dependent on the $isospin$ space $(T,T_{z})$
as
\ba
\left.
\BA{c}
\rho^{T,T_{z}}_{k}
\!\equiv\!
{\displaystyle \frac{\sqrt 2}{\sqrt A}} \!
\sum_{p, \tau_{z},\tau^{\prime}_{z}} \!
\langle
{\displaystyle \frac{1}{2}} \tau_{z}
{\displaystyle \frac{1}{2}} \tau^{\prime}_{z} 
|T T_{z}
\rangle
a^{\dag}_{p+\frac{k}{2}, \tau_{z}}
(-1)^{\frac{1}{2} \!+\! \tau^{\prime}_{z}}
a_{p-\frac{k}{2}, -\tau^{\prime}_{z}} , ~~
{\rho^{T,T_{z}}_{k}}^{\dag}
\!=\!
(-1)^{T_{z}} \rho^{T,-T_{z}}_{-k} , \\
\\[-8pt]
\rho^{0,0}_{0}
\!=\!
{\displaystyle \frac{\sqrt 2}{\sqrt{2A}}}
\sum_{p,\tau_{z}} \!
a^{\dag}_{p,\tau_{z}} a_{p,\tau_{z}}
\!=\!
{\displaystyle \frac{1}{\sqrt{A}}} (N \!+\! Z)
\!=\!
\sqrt A , \\
\\[-8pt]
\rho^{1,0}_{0}
\!=\!
{\displaystyle \frac{\sqrt 2}{\sqrt{2A}}}
\sum_{p} \!
\left( \!
a^{\dag}_{p,N} a_{p,N}
\!-\!
a^{\dag}_{p,P} a_{p,P} \!
\right)
\!=\!
{\displaystyle \frac{1}{\sqrt{A}}} (N \!-\! Z) ,\\
\\[-8pt]
\rho^{1,1}_{0}
\!=\!
- {\displaystyle \frac{\sqrt 2}{{\sqrt A}}}
\sum_{p} \!
a^{\dag}_{p,N} a_{p,P} ,~
\rho^{1,-1}_{0}
\!=\!
{\displaystyle \frac{\sqrt 2}{{\sqrt A}}}
\sum_{p} \!
a^{\dag}_{p,P} a_{p,N} ,
\EA
\right\}
\label{FcomponentDensityOp}
\ea
where $N, ~Z$ and $A$ are the total numbers of the neutron, the proton and
the N-P System under consideration, respectively
\cite{Lipkin.65a,Lipkin.65b,RoweWood.10}.
Substituting
(\ref{FexpaFieldop})
and
(\ref{FexpaIntPot})
into the expression for the interaction $V$ in
(\ref{Hamiltonian}),
the $V$ is expressed as
\ba
\BA{c}
V
\!=\!
{\displaystyle \frac{1}{2L}}
\sum_{\{k\},\{\tau_{z}\},T,T_{z}} \!
\nu_{T}(k)
\langle
{\displaystyle \frac{1}{2}} \tau^{\prime}_{z}
{\displaystyle \frac{1}{2}} \tau^{\prime \prime}_{z}
|T T_{z}
\rangle
\langle
{\displaystyle \frac{1}{2}} \tau^{\prime \prime \prime}_{z}
{\displaystyle \frac{1}{2}} \tau^{\prime \prime \prime \prime}_{z}
|T T_{z}
\rangle
a^{\dag}_{k^{\prime} \tau^{\prime}_{z}}
a^{\dag}_{k^{\prime \prime} \tau^{\prime \prime}_{z}}
a_{k^{\prime \prime} + k, \tau^{\prime \prime \prime}_{z}}
a_{k^{\prime} - k, \tau^{\prime \prime \prime \prime}_{z}} \\
\\[-8pt]
\!=\!
{\displaystyle \frac{A}{4L}} \!\!
\sum_{T,T_{z},k} \!
\left\{ \!
\sum_{T^{\prime}}
(2T^{\prime} \!\!+\!\! 1) 
W \!\!
\left( \!
{\displaystyle \frac{1}{2}}
{\displaystyle \frac{1}{2}}
{\displaystyle \frac{1}{2}}
{\displaystyle \frac{1}{2}};\!
T T^{\prime} \!
\right) \!
\nu_{T^{\prime}}(k) \!
\right\} \!
\rho^{T,T_{z}}_{k} \!
(\!-\!1)^{T_{z}} \! 
\rho^{T, -T_{z}}_{-k}
\!\!-\!\!
{\displaystyle \frac{A}{4L}} \!\!
\sum_{T,k}
(2T \!\!+\!\! 1) 
\nu_{T}(k) . 
\EA
\label{Interaction}
\ea
Here we have used the relations
(\ref{ACR})
and
(\ref{FcomponentDensityOp}).
Then the final expression for the Hamiltonian $H$ is given as
\ba
\left.
\BA{c}
H
\!=\!
T \!+\! V
\!=\!
\sqrt{2}
\sum_{k,\tau_{z}} \!
{\displaystyle \frac{\hbar^{2}k^{2}}{2m}}
(-1)^{\frac{1}{2} \!-\! \tau_{z}}
\langle
{\displaystyle \frac{1}{2}} \tau_{z}
{\displaystyle \frac{1}{2}} -\tau_{z}
|0 0
\rangle
a^{\dag}_{k \tau_{z}}
a_{k \tau_{z}} \\
\\[-8pt]
+
{\displaystyle \frac{A}{4L}} 
\sum_{T,T_{z},k}
\nu^{F}_{T}(k)
\rho^{T, T_{z}}_{k}
(-1)^{T_{z}} \! 
\rho^{T, -T_{z}}_{-k}
\!-\!
{\displaystyle \frac{A}{4L}} 
\sum_{T,k}
(2T \!+\! 1) 
\nu_{T}(k) ,  \\
\\[-8pt]
\nu^{F}_{T}(k)
\!\equiv\!
\sum_{T^{\prime}}
(2T^{\prime} \!\!+\!\! 1) 
W \!
\left( \!
{\displaystyle \frac{1}{2}}
{\displaystyle \frac{1}{2}}
{\displaystyle \frac{1}{2}}
{\displaystyle \frac{1}{2}};
T T^{\prime} \!
\right) \!
\nu_{T^{\prime}}(k) . 
\EA
\right\}
\label{ExpressionforHamiltonian}
\ea
Using
(\ref{ACR}),
the commutation relation between
$\rho^{T_{1},T_{z1}}_{k_{1}}$
and
$\rho^{T_{2},T_{z2}}_{k_{2}}$
is calculated as
\ba
\BA{c}
\left[ \rho^{T_{1},T_{z1}}_{k_{1}}, \rho^{T_{2},T_{z2}}_{k_{2}} \right]
\!=\!
{\displaystyle \frac{\sqrt 2}{\sqrt A}}
\sum_{T_{3},T_{z3}}
\left\{ (-1)^{T_{1} \!+\! T_{2} \!+\! T_{3}} \!-\! 1 \right\} \\
\\[-8pt]
\times
\sqrt{ (2T_{1} \!+\! 1) (2T_{2} \!+\! 1)}
W \!
\left( \!
{\displaystyle \frac{1}{2}} {\displaystyle \frac{1}{2}} T_{1} T_{2};
T_{3} {\displaystyle \frac{1}{2}} \!
\right) \!
\langle
T_{1} T_{z1}
T_{2} T_{z2}
|T_{3} T_{z3}
\rangle
\rho^{T_{3},T_{z3}}_{k_{1} \!+\! k_{2}} ,
\EA
\label{CRrho}
\ea
detailed calculations of which are given in Appendix A.
With the aid of 
(\ref{ExpressionforHamiltonian})
and
(\ref{CRrho}),
the commutation relation between $V$ and $\rho^{T,T_{z}}_{k}$
is computed as
\ba
\BA{c}
\left[ V, \rho^{T,T_{z}}_{k} \right]
\!=\!
{\displaystyle \frac{\sqrt A}{2 \sqrt 2 L}} 
\sum_{T_{1},T_{z1},k_{1}}
\nu^{F}_{T_{1}}(k_{1})
\sum_{T_{2},T_{z2}} \!\!
\left\{ (-1)^{T \!+\! T_{1} \!+\! T_{2}} \!-\! 1 \right\} \!\!
\sqrt{ (2T_{1} \!+\! 1) (2T_{2} \!+\! 1)} \\
\\[-8pt]
\times
W \!
\left( \!
{\displaystyle \frac{1}{2}} {\displaystyle \frac{1}{2}} T_{1} T_{2};
T {\displaystyle \frac{1}{2}} \!
\right) \!
\langle
T_{1} T_{z1}
T_{2} T_{z2}
|T T_{z}
\rangle \!
\left\{
\rho^{T_{1}, T_{z1}}_{k_{1}} \!
\rho^{T_{2},T_{z2}}_{k \!-\! k_{1}} 
\!+\!
\rho^{T_{2},T_{z2}}_{k \!+\! k_{1}} 
\rho^{T_{1}, T_{z1}}_{- k_{1}} \!
\right\} .
\EA
\label{CRVrho}
\ea

In the definition of the Fourier component of the density operator
(\ref{FcomponentDensityOp}),
we concentrate on the two following cases

\vspace{0.5cm}

(i) : $T \!=\! 0,~T_{z} \!=\! 0$,\\
\\[-18pt]

(ii): $T \!=\! 1,~T_{z} \!=\! 0$.

\vspace{0.5cm}

Then, we have
\ba
\!\!\!\!\!\!\!\!
\left.
\BA{c}
(\mbox{i}): \rho^{0,0}_{k}
\!\equiv\!
{\displaystyle \frac{\sqrt 2}{\sqrt A}} \!
\sum_{p, \tau_{z}} \!
\langle
{\displaystyle \frac{1}{2}} \tau_{z}
{\displaystyle \frac{1}{2}} \!-\!\tau_{z}
|0 0
\rangle
a^{\dag}_{p+\frac{k}{2}, \tau_{z}} \!
(\!-\!1)^{\frac{1}{2} \!-\! \tau_{z}} \!
a_{p-\frac{k}{2}, \tau_{z}}
\!\!=\!\!
{\displaystyle \frac{1}{\sqrt A}} \!
\sum_{p, \tau_{z}} \!
a^{\dag}_{p+\frac{k}{2}, \tau_{z}} \!
a_{p-\frac{k}{2}, \tau_{z}}, \\
\\[-8pt]
{\rho^{0,0}_{k}}^{\dag}
\!=\!
\rho^{0,0}_{-k} ,~~
\left[ \rho^{0,0}_{k}, \rho^{0,0}_{k^{\prime}} \right]
\!=\!
0 , ~~
\left[V, \rho^{0,0}_{k} \right]
\!=\!
0 ,\\
\\[-8pt]
(\mbox{ii}): \rho^{1,0}_{k}
\!\equiv\!
{\displaystyle \frac{\sqrt 2}{\sqrt A}} \!
\sum_{p, \tau_{z}} \!
\langle
{\displaystyle \frac{1}{2}} \tau_{z}
{\displaystyle \frac{1}{2}} \!-\!\tau_{z}
|1 0
\rangle
a^{\dag}_{p+\frac{k}{2}, \tau_{z}} \!
(\!-\!1)^{\frac{1}{2} \!-\! \tau_{z}} \!
a_{p-\frac{k}{2}, \tau_{z}} 
\!\!=\!\!
{\displaystyle \frac{2}{\sqrt A}} \!
\sum_{p, \tau_{z}} \!
\tau_{z}
a^{\dag}_{p+\frac{k}{2}, \tau_{z}} \!
a_{p-\frac{k}{2}, \tau_{z}} , \\
\\[-8pt]
{\rho^{1,0}_{k}}^{\dag}
\!=\!
\rho^{1,0}_{-k} , ~~
\left[ \rho^{1,0}_{k}, \rho^{1,0}_{k^{\prime}} \right]
\!=\!
0 , ~~
\left[V, \rho^{1,0}_{k} \right]
\!=\!
0 .
\EA \!\!
\right\}
\label{FcomponentDensityOpT0T1}
\ea
It is possible to prove the last relation in the case (ii) of
(\ref{FcomponentDensityOpT0T1}).
$\!\!$Using
(\ref{CRVrho}),
the commutator
$\left[V, \rho^{1,0}_{k} \right]$
is calculated as\\[-16pt]
\ba
\BA{l}
\left[V, \rho^{1,0}_{k} \right]
\!=\!
-{\displaystyle \frac{\sqrt A}{\sqrt 2 L}} \!
\sum_{k_{1}} 
\nu^{F}_{T_{1}}(k_{1}) \!
\left\{
\rho^{1, 1}_{k_{1}}
\! 
\rho^{1,-1}_{k \!-\! k_{1}} 
\!+\!
\rho^{1,-1}_{k \!+\! k_{1}}
\! 
\rho^{1, 1}_{- k_{1}}
\!-\!
\rho^{1, -1}_{k_{1}}
\!
\rho^{1,1}_{k \!-\! k_{1}} 
\!-\!
\rho^{1,1}_{k \!+\! k_{1}}
\! 
\rho^{1, -1}_{- k_{1}}
\right\} ,
\EA
\ea\\[-10pt]
which vanishes
if the condition 
$
\nu^{F}_{T_{1}} \! (\!k_{1}\!)
\!\!=\!\!
\nu^{F}_{T_{1}} \! (\!k \!+\! k_{1}\!)
$
is satisfied.
This means 
$\nu^{F}_{T_{1}} \! (\!k_{1}\!)$
is constant.
Then the variables $\rho^{0,0}_{k}$ and $\rho^{1,0}_{k}$ become
good collective variables.
From now on we will find ${exact}$ canonically conjugate momenta
to these collective variables. 
Following Tomonaga, first we introduce collective 
momenta associated with the $\rho^{0,0}_{-k}$ through\\[-14pt] 
\ba
{\pi_{k}^{0,0}}
=
{\displaystyle \frac{m} {{k}^{2}}}
\dot{\left(\rho_{-k}^{0,0}\right)}
=
{\displaystyle \frac{m} {{k}^{2}}}
{\displaystyle \frac{i} {\hbar}}
[H, \rho^{0,0}_{-k} ]
=
{\displaystyle \frac{m} {{k}^{2}}}
{\displaystyle \frac{i} {\hbar}}
[T, \rho^{0,0}_{-k} ]
=
{\pi_{-k}^{0,0}}^{\dag }  ,
~(k \!\ne\! 0) ~.
\label{pik} 
\ea
Calculating the commutator (\ref{pik}),
we obtain explicit 
expressions for the associated collective variables
${\pi_{k}^{0,0}}$
as\\[-14pt]
\ba
\!\!\!\!
{\pi_{k}^{0,0}}
\!=\!
-{\displaystyle \frac{i \sqrt 2 \hbar}{\sqrt A {k}^{2}}} \!
\sum_{p, \tau_{z}} 
pk
\langle
{\displaystyle \frac{1}{2}} \tau_{z}
{\displaystyle \frac{1}{2}} \!-\!\tau_{z}
|0 0
\rangle
a^{\dag}_{p-\frac{k}{2}, \tau_{z}} \!
(\!-\!1)^{\frac{1}{2} \!-\! \tau_{z}} \!
a_{p+\frac{k}{2}, \tau_{z}}
\!\!=\!\!
-{\displaystyle \frac{i \hbar}{\sqrt A k}} \!
\sum_{p, \tau_{z}} 
p
a^{\dag}_{p-\frac{k}{2}, \tau_{z}} \!
a_{p+\frac{k}{2}, \tau_{z}} , 
\label{pik2}  
\ea\\[-10pt]
where we have used the explicit expression for
the kinetic operator $T$ in
(\ref{ExpressionforHamiltonian})
and
\ba
[T, \rho^{0,0}_{-k} ]
\!=\!
{\displaystyle \frac{\sqrt 2}{\sqrt A}}
{\displaystyle \frac{\hbar^{2}}{2m}} 
\sum_{p, \tau_{z}}
\left\{ \!
\left( \! p \!-\! {\displaystyle \frac{k}{2}} \right)^{\!2}
\!\!-\!\!
\left( \! p \!+\! {\displaystyle \frac{k}{2}} \right)^{\!2} \!
\right\} \!
\langle
{\displaystyle \frac{1}{2}} \tau_{z}
{\displaystyle \frac{1}{2}} \!-\!\tau_{z}
|0 0
\rangle
a^{\dag}_{p-\frac{k}{2}, \tau_{z}} \!
(\!-\!1)^{\frac{1}{2} \!-\! \tau_{z}} \!
a_{p+\frac{k}{2}, \tau_{z}} .
\label{commuTrho}  
\ea\\[-10pt] 
At first, this $\pi^{0,0}_{k}$ is regarded as 
the collective momentum conjugate to
the density operator $\rho^{0,0}_{k}$
in the sense of Tomonaga 
\cite{Tomo.55a,Tomo.55b}. 
Unfortunately, however, 
the commutation relations among the variables 
$\rho^{0,0}_{k}$ and $\pi^{0,0}_{k}$,
lead to a following result:\\[-10pt]
\ba
\left.
\BA{ll}
&[\rho^{0,0}_{k} ,\rho^{0,0}_{k'}]
=
0, \\
\\[-8pt]
&[\pi^{0,0}_{k} ,\rho^{0,0}_{k'}] 
=
- {\displaystyle \frac{i\hbar }{\sqrt A}}
{\displaystyle \frac{k'}{k}}
\rho^{0,0}_{k'-k} , \\
\\[-8pt]
&[\pi^{0,0}_{k} ,\pi^{0,0}_{k'}]
=
- {\displaystyle \frac{i\hbar }{\sqrt{A} kk'}}
({k}^{2} - {k'}^{2})
\pi^{0,0}_{k+k'} .
\EA
\right\}
\label{CRpirho}  
\ea
These commutation relations have quite the same structures as those of the 
commutation relations
obtained at the first stage in the
Sunakawa's discrete integral equation method
\cite{SYN.62}.
As is shown from
(\ref{CRpirho}), 
the right-hand side (RHS) of the second line does not take the value
${-\ i\hbar{\delta }_{kk'}}$ 
and the third one does not vanish. Detailed calculations for them are given 
in Appendix A.
Then from these facts, it is self-evident that the 
variables $\rho^{0,0}_{k'}$ and $\pi^{0,0}_{k}$
are not canonically conjugate to each 
other if we take into account contributions of the order of 
$\frac{1}{\sqrt A}$.

\vspace{0.3cm} 

For $T \!=\! 1~\mbox{and}~T_{z} \!=\! 0$,
we have the following commutation relations:
\ba
\!\!\!\!
\left.
\BA{ll}
&[\rho^{1,0}_{k} ,\rho^{1,0}_{k'}]
=
0, ~~
[\rho^{1,0}_{k} ,\rho^{0,0}_{k'}]
=
0, \\
\\[-6pt]
&[\pi^{1,0}_{k} ,\rho^{1,0}_{k'}] 
=
- {\displaystyle \frac{i\hbar }{\sqrt A}}
{\displaystyle \frac{k'}{k}}
\rho^{0,0}_{k'-k} , ~~
[\pi^{1,0}_{k} ,\rho^{0,0}_{k'}] 
=
- {\displaystyle \frac{i\hbar }{\sqrt A}}
{\displaystyle \frac{k'}{k}}
\rho^{1,0}_{k'-k} , \\
\\[-10pt]
&[\pi^{1,0}_{k} ,\pi^{1,0}_{k'}]
=
- {\displaystyle \frac{i\hbar }{\sqrt{A} kk'}}
({k}^{2} - {k'}^{2})
\pi^{0,0}_{k+k'} ,~~
[\pi^{1,0}_{k} ,\pi^{0,0}_{k'}]
=
- {\displaystyle \frac{i\hbar }{\sqrt{A} kk'}}
({k}^{2} - {k'}^{2})
\pi^{1,0}_{k+k'} .
\EA
\right\}
\label{CRpirho2}  
\ea
As is clear the structures of the commutators
(\ref{CRpirho2}),
they are shown to have the twisted property
in the {\it isospin} space $(T,T_z)$,
comparing with those of  the commutators
(\ref{CRpirho}).
This is a quite different behavior from the behavior
for an {\it isospin}-less Fermion system.


\newpage

\def\thesection{\arabic{section}}
\setcounter{equation}{0}
\renewcommand{\theequation}{\arabic{section}.\arabic{equation}}     
\section{Exact Canonically Conjugate Momenta}

In order to overcome the difficulties mentioned in the preceding section, 
we define the ${exact}$ canonically conjugate momenta
$\Pi^{0,0}_{k}$ by\\[-16pt]
\ba
\BA{c}
\Pi^{0,0}_{k}
\!=\!
\pi^{0,0}_{k} 
\!-\!
{\displaystyle \frac{1}{ \sqrt {A}k}}
\sum_{p \ne k} p
\rho^{0,0}_{p-k} \Pi^{0,0}_{p}~ 
(k \!\ne\!  0) ,~~
\Pi^{0,0}_{k}{}^{\dag }
\!=\!
\Pi^{0,0}_{k}
\EA .
\label{exactPi}  
\ea\\[-14pt]
This type of the discrete integral equation was first presented
in the Sunakawa's second quantized collective formalism
for an interacting Bose system 
\cite{SYN.62}. 
It was also proposed by us and one of the present author's (S.N.)
in the first quantized manner for a description of
a quadrupole type nuclear collective motion 
\cite{NishProvi.14}.
As is clear from the structure of
(\ref{exactPi}), 
the variables 
$\Pi^{0,0}_{k}$ are no longer one-body operators but essentially many-body 
operators.
From (\ref{exactPi}),
we get the ${exact}$ canonical 
commutation relations\\[-16pt] 
\ba
[\rho^{0,0}_{k} , \rho^{0,0}_{k'}]
\!=\! 
0 ,~~
[\Pi^{0,0}_{k}, \rho^{0,0}_{k'}]
\!=\! 
- i\hbar {\delta }_{kk'} ,~~
[\Pi^{0,0}_{k} , \Pi^{0,0}_{k'}] = 0  ,
\label{exactCRs}  
\ea
and the commutators among
the new $\Pi^{0,0}_{k}$ and 
the old $\pi^{0,0}_{k}$
as\\[-16pt]
\ba
[\pi^{0,0}_{k} , \Pi^{0,0}_{k'}]
\!=\!
{\displaystyle \frac{i\hbar}{ \sqrt {A}k}} (k+k')
\Pi^{0,0}_{k+k'}  ,
\label{CRpiPi}  
\ea\\[-16pt]
derivation of
(\ref{CRpiPi})
is given in Appendix B.
Following Sunakawa's method,
the above 
${exact}$ canonical commutation relations are proved also in Appendix B.
Thus, we have proved the ${exact}$ canonical commutation relations for 
${\rho^{0,0}_{k}}$ and ${\Pi^{0,0}_{k}}$. 
The hermiticity property
${\Pi^{0,0 \dag}_{k}}
\!=\! 
{\Pi^{0,0}_{-k}}$
can be proved with the help of 
(\ref{FcomponentDensityOpT0T1}), (\ref{pik}) and 
(\ref{CRPirho}).

For $T \!=\! 1~\mbox{and}~T_{z} \!=\! 0$,
we have the following commutation relations:\\[-10pt]
\ba
\BA{c}
\Pi^{1,0}_{k}
\!=\!
\pi^{1,0}_{k} 
\!-\!
{\displaystyle \frac{1}{ \sqrt {A}k}}
\sum_{p \ne k} p
\rho^{0,0}_{p-k} \Pi^{1,0}_{p}~ 
(k \!\ne\!  0) ,~~
\Pi^{1,0}_{k}{}^{\dag }
\!=\!
\Pi^{1,0}_{-k} ,
\EA 
\label{exactPi2}  
\ea
\vspace{-0.3cm}
\ba
[\rho^{1,0}_{k} , \rho^{1,0}_{k'}]
\!=\! 
0 ,~~
[\Pi^{1,0}_{k}, \rho^{1,0}_{k'}]
\!=\! 
- i\hbar {\delta }_{kk'} .
\label{exactCRs2}  
\ea
As shown before,
the structures of the commutators among
$\rho^{0,0}_{k}, \rho^{1,0}_{k}, \pi^{0,0}_{k}
~\mbox{and}~\rho^{1,0}_{k}$ in
(\ref{CRpirho2})
have the twisted property
in the {\it isospin} space $(T,T_z)$.
Due to this twisted property,  
unfortunately, 
the commutators
$[\Pi^{1,0}_{k} , \Pi^{1,0}_{k'}]$
do not vanish.
Then,
strictly speaking,
the $\rho^{1,0}_{k}$ and $\Pi^{1,0}_{k}$ are not
the $exact$ canonical variables
with each other.
Hereafter we concentrate on the first case
(i) : $T \!=\! 0$ and $T_{z} \!=\! 0$
given in
(\ref{FcomponentDensityOpT0T1}).


\newpage

\def\thesection{\arabic{section}}
\setcounter{equation}{0}
\renewcommand{\theequation}{\arabic{section}.\arabic{equation}}     
\section{The ${\Pi_{k}}$- and ${\rho_{k}}$-Dependence of the Hamiltonian}
From the original Hamiltonian
(\ref{Hamiltonian})
we derive here a collective 
Hamiltonian in terms of the ${exact}$ canonical variables
${\rho^{0,0}_{k}}$ and ${\Pi^{0,0}_{k}}$. 
The potential part ${V}$ is already written in terms of ${\rho^{0,0}_{k}}$. 
Our task is therefore to express the kinetic part ${T}$ in terms of the 
${exact}$ canonical variables
${\rho^{0,0}_{k}}$ and ${\Pi^{0,0}_{k}}$.
For this 
purpose, following Sunakawa's method, first we expand it in a 
power series of the ${exact}$ canonically conjugate momenta
${\Pi^{0,0}_{k}}$ 
as follows: 
\ba
\BA{c}
T
\!=\! 
{T}_{0}(\rho)
\!+\! 
\sum_{p \ne 0}
{T}_{1} (\rho ; p  ) \Pi^{0,0}_{p}
\!+\!
\sum_{p \ne 0,\ q \ne 0}
{T}_{2} (\rho ; p,q )
\Pi^{0,0}_{p}\Pi^{0,0}_{q}\ 
\!+\! 
\cdot \cdot \cdot  ,
\EA
\label{Texpansion}  
\ea
where\\[-26pt]
\ba
{T}_{2} (\rho ; p,q )
\!=\! 
{T}_{2} 
( \rho ; q,p ) .
\ea
In Eq.
(\ref{Texpansion}), 
${T_{n} (n \neq 0)}$ are the unknown expansion coefficients. 
In order to get their explicit expressions, we take the commutators 
between ${T}$ and ${\rho^{0,0}_{k}}$.
On the other hand, from  
(\ref{FcomponentDensityOpT0T1}) and (\ref{pik}),
we can calculate directly values of the commutators 
between ${T}$ and ${\rho^{0,0}_{k}}$ by using the definition (\ref{exactPi}) 
as follows:\\[-14pt]
\ba
\BA{lll}
[T , \rho^{0,0}_{-k}]
\!\!&=&\!\!
{\displaystyle {\frac{ \hbar }{i}}}
\dot{ \left( {\rho}^{0,0}_{k} \right) }
\!=\!
-{\displaystyle \frac{i \hbar {k}^{2}}{m}}
\pi^{0,0}_{-k} \\
\\[-10pt]
\!\!&=&\!\!
-{\displaystyle \frac{i \hbar {k}^{2}}{m}}
\Pi^{0,0}_{-k} 
\!+\!
{\displaystyle {\frac{i\hbar k}{m \sqrt {A}}}}
\sum_{p \ne -k} p
\rho^{0,0}_{p+k} \Pi^{0,0}_{p}  .
\EA
\label{CRTrho}  
\ea
Using Eq. (\ref{CRPirho}) successively, we can easily obtain the commutators 
\ba
[[T, \rho^{0,0}_{k}] , \rho^{0,0}_{k'}]
\!=\!
\left\{ \!\!\!\!\!\!
\BA{cc}
&-{\displaystyle \frac{ \hbar^{2} {k}^{2}}{m}} ~,~  \mbox{for}~~k' \!=\! -k ,\\
\\[-8pt]
&{\displaystyle \frac{{\hbar}^{2}kk'}{m \sqrt {A}}}
\rho^{0,0}_{k+k'} ~,~  \mbox{for}~~k'  \!\ne\!  - k ,
\EA
\right.
\label{CRTrhorho}
\ea
\ba
[[[T , \rho^{0,0}_{k} ] , \rho^{0,0}_{ k'}] , \rho^{0,0}_{k''}]
\!=\!
0 ,
\label{CRTrhorhorho} 
\ea
and so on. Comparing the above results with the commutators between ${T}$ 
of
(\ref{Texpansion}) and $\phi^{0,0}_{k}$,
we can determine the coefficients 
${T_{n} (n \!\neq\! 0)}$. Then we can express the kinetic part ${T}$ in terms 
of the ${exact}$ canonical variables 
$\rho^{0,0}_{k}$ and $\Pi^{0,0}_{k}$ as follows:
\ba
\BA{c}
T
\!=\!
{T}_{0}(\rho)
\!+\!
{\displaystyle \frac{1}{2m}}
\sum_{k}
{k}^{2}
\Pi^{0,0}_{k} \Pi^{0,0}_{-k}
\!-\! 
{\displaystyle \frac{1}{2m \sqrt {A}}}
\sum_{p \ne 0, q \ne 0, p+q \ne 0} 
pq
\rho^{0,0}_{p+q} \Pi^{0,0}_{p} \Pi^{0,0}_{q} .
\EA
\label{exactTPi}  
\ea
Here, we should stress that up to the present stage, all the expressions 
are derived without any approximation. 

Our remaining task in this section is to determine the term
$T_{0}(\rho)$ in 
(\ref{Texpansion})
which depends only on $\rho^{0,0}_{k}$. 
For this purpose, we also expand it in a power series of the collective 
coordinates $\rho^{0,0}_{k}$ in the form 
\ba
\BA{c}
{T}_{0} (\rho)
\!=\! 
{C}_{0}
\!+\!
\sum_{p \ne 0} 
{C}_{1}(p) \rho^{0,0}_{p}
\!+\!
\sum_{p \ne 0,\ q \ne 0}
{C}_{2} (p,q)
\rho^{0,0}_{p} \rho^{0,0}_{q} 
\!+\! 
\cdot \cdot \cdot  ,
\EA
\label{T0rho}  
\ea
where
${C}_{2} (p,q ) \!=\!  {C}_{2} (q,p )$.
The expansion coefficients should be determined by a
procedure similar to the one used in the previous section.
From the definition 
(\ref{exactPi}),
we get easily the discrete integral equation
\ba
\BA{c}
[\Pi^{0,0}_{k}  , {T}_{0}(\rho)]
\ =\
{f}_{k} (\rho  )\ 
\!-\!
{\displaystyle \frac{1}{\sqrt {A}k}}
\sum_{p \ne 0, k}
p
\rho^{0,0}_{p-k}
[\Pi^{0,0}_{p}  , {T}_{0}(\rho)]  ,
\EA
\label{CRPiT0}  
\ea
and the inhomogeneous term ${{f}_{k}(\rho)}$ becomes\\[-16pt]
\ba
\!\!\!\!
\BA{lll}
{f}_{k} ( \rho )
\!\!\!\!&\equiv&\!\!\!\! 
[\pi^{0,0}_{k}  , {T}_{0}(\rho)]
\!-\!
\left[
\pi^{0,0}_{k} , T 
\!-\!
{\displaystyle \frac{1}{2m}} \!
\sum_{p} \! {p}^{2} 
\Pi^{0,0}_{p} \Pi^{0,0}_{-p} 
\!+\! 
{\displaystyle \frac{1}{ 2m \sqrt {\!A}}} \!
\sum_{p \ne 0, q \ne 0, p+q \ne 0} 
pq
\rho^{0,0}_{p+q}
\Pi^{0,0}_{p}  \Pi^{0,0}_{q} 
\right] \\
\\[-10pt]
\!\!&=&\!\!
[\pi^{0,0}_{k} , T]
\!-\!
{\displaystyle \frac{i\hbar}{mA}} 
\sum_{p \ne 0,\ q \ne 0} 
pq
\rho^{0,0}_{p+q-k}
\Pi^{0,0}_{p} \Pi^{0,0}_{q} ,~ (k \!\ne\! 0)
\EA
\label{fkrho}  
\ea
with the aid of the result (\ref{exactTPi}) and the commutation relation 
(\ref{CRpiPi}).
At a first glance, the operator-valued function 
$f_{k}(\rho)$ seems to be dependent on $\Pi^{0,0}_{k}$.
However, it turns out that
$f_{k}(\rho)$
does not really depend on 
$ \Pi^{0,0}_{k}$
because it commutes with $\rho^{0,0}_{k}$ as shown below\\[-14pt]
\ba
\BA{lll}
[\rho^{0,0}_{k'} , {f}_{k}(\rho)]
\!\!&=&\!\!
[\rho^{0,0}_{k'} , [\pi^{0,0}_{k} , T]]
\!+\!
{\displaystyle \frac{2{\hbar}^{2}k'}{mA}}
\sum_{p \ne 0} 
p
\rho^{0,0}_{p-(k-k')} \Pi^{0,0}_{p} \\
\\[-10pt]
\!\!&=&\!\!
- 2{\hbar}^{2}
{\displaystyle \frac{k'(k-k')}{m \sqrt {A}}}
\pi^{0,0}_{k-k'}  
\!+\!
{\displaystyle \frac{2{\hbar}^{2}k'}{mA}}
\sum_{p \ne 0}
p
\rho^{0,0}_{p-(k-k')}
\Pi^{0,0}_{p} \\
\\[-10pt]
\!\!&=&\!\! 
- 2{\hbar}^{2}
{\displaystyle \frac{k'(k-k')}{m \sqrt {A}}}
\left\{ 
\pi^{0,0}_{k-k'} 
\!-\! 
{\displaystyle \frac{1}{\sqrt {A}(k-k')}}
\sum_{ p~all}
p
\rho^{0,0}_{p-(k-k')} \Pi^{0,0}_{p} 
\right\}
\!=\!
0 .
\EA
\label{CRrhofk}  
\ea
In the above we have used the commutation relation
\ba
\BA{c}
[\rho^{0,0}_{k'} , [\pi^{0,0}_{k} , T]]
\!=\! 
-
2{\hbar}^{2}
{\displaystyle \frac{k'(k-k')}{m \sqrt {A}}}
\pi^{0,0}_{k-k'}   ,
\EA
\label{CRrhopiT}  
\ea\\[-12pt]
which is proved with the help of the commutation relation between
$\pi^{0,0}_{k}$ and ${T}$,\\[-12pt]
\ba
\!\!\!\!\!\!
[\pi^{0,0}_{k} , T]
\!=\!
-{\displaystyle \frac{i \hbar}{\sqrt A}}
{\displaystyle \frac{\hbar^{2}}{2m}} 
{\displaystyle \frac{1}{k}} \!\!
\sum_{p, \tau_{z}} \!
p \!
\left\{ \!\!
\left( \! p \!+\! {\displaystyle \frac{k}{2}} \! \right)^{\!\!2}
\!\!-\!\!
\left( \! p \!-\! {\displaystyle \frac{k}{2}} \! \right)^{\!\!2} \!\!
\right\} \!
a^{\dag}_{p-\frac{k}{2}, \tau_{z}} \!
a_{p+\frac{k}{2}, \tau_{z}}
\!\!=\!\!
-{\displaystyle \frac{i \hbar^{3}}{ m \! \sqrt {\!A}}} \!
\sum_{p, \tau_{z}}
p^{2} \!
a^{\dag}_{p-\frac{k}{2}, \tau_{z}} \!
a_{p+\frac{k}{2}, \tau_{z}} .
\label{commupiT}  
\ea\\[-8pt] 
Up to the present stage,
all the expressions are exact.

From now on, we make an approximation to calculate $T_{0}(\rho)$ 
up to the order of $\frac{1}{A}$. 
First we use $\rho^{0,0}_{0} \!=\! \sqrt {A}$ and
$\Pi^{0,0}_{k} \!\cong\! \pi^{0,0}_{k}$ and give
approximate expressions for
$\pi^{0,0}_{k}$ and  $\rho^{0,0}_{k}$
as\\[-12pt]
\ba
\BA{c}
\pi^{0,0}_{k}
\!\cong\!
- {\displaystyle \frac{i\hbar}{2}} 
\sum_{\tau_{z}} \!
\left( 
\overline{\theta}a_{k, \tau_{z}}
\!-\!
a^{\dag}_{-k, \tau_{z}}\theta 
\right) ,~
\rho^{0,0}_{k}
\!\cong\!
\sum_{\tau_{z}} \!
\left( 
\overline{\theta}a_{-k, \tau_{z}}
\!+\!
a^{\dag}_{k, \tau_{z}}\theta 
\right) , 
\EA
\label{approxpirho}  
\ea\\[-12pt]
and we regard the operators
$a_{0, \tau_{z}}$ and  $a^{\dag}_{0, \tau_{z}}$
as $c$-numbers but with the Grassmann variables\\[-12pt]
\ba
a_{0, \tau_{z}}
\!\cong\!
\sqrt{A} ~\! \theta ,~
a^{\dag}_{0, \tau_{z}}
\!\cong\!
\sqrt{A} ~\! \overline{\theta} ,
\label{approxa0}  
\ea\\[-14pt]
where the 
$\theta$ and $\overline{\theta}$
are Grassmann variables and anti-commute with
$a_{k, \tau_{z}}$ and $a^{\dag}_{k, \tau_{z}}$
\cite{Berezin.66,Casalbuoni.76a,Casalbuoni.76b}.
These  Grassmann variables play crucial roles to estimate
the inhomogeneous term ${{f}_{k}(\rho)}$
(\ref{fkrho}),
though they were unnecessary to compute the
Sunakawa's
${{f}_{k}(\rho)}$ 
in the case of a Bose system
\cite{SYN.62}.
Then the second term in the last line of 
(\ref{fkrho})
is approximately computed as\\[-12pt]
\ba
\!\!\!\!
\BA{c}
-
{\displaystyle \frac{i\hbar}{mA}} 
\sum_{p \ne 0,\ q \ne 0} 
p.q.
\rho^{0,0}_{p+q-k}
\Pi^{0,0}_{p} \Pi^{0,0}_{q} 
\!\cong\!
-
{\displaystyle \frac{i\hbar}{m \sqrt{A}}} 
\sum_{p \ne 0, p \ne k} 
p(k\!-\!p)
\pi^{0,0}_{p} \pi^{0,0}_{k-p} \\
\\[-10pt]
\!=\!
{\displaystyle \frac{i\hbar^{3}}{4 m \sqrt{A}}} \!
\sum_{p \ne 0, p \ne k}
p(k\!-\!p) \!
\sum_{\tau_{z}} \!
\left( 
\overline{\theta}a_{p, \tau_{z}} 
\!-\!
a^{\dag}_{-p, \tau_{z}}\theta 
\right) \!
\sum_{\tau'_{z}} \!
\left( 
\overline{\theta}a_{k-p, \tau'_{z}} 
\!-\!
a^{\dag}_{-(k-p), \tau'_{z}}\theta 
\right)
\!+\! 
O \left( \! {\displaystyle \frac{1}{A}} \! \right)  \\
\\[-10pt]
\!=\!
{\displaystyle \frac{i\hbar^{3}}{4 m \sqrt{A}}} \!
\sum_{p \ne 0, p \ne k}
p(k\!-\!p) \!
\left( 
\rho^{0,0}_{-p} 
\!-\!
2 \! \sum_{\tau_{z}} \! a^{\dag}_{-p, \tau_{z}}\theta 
\right) \!
\left( 
\rho^{0,0}_{-(k-p)} 
\!-\!
2 \! \sum_{\tau'_{z}} \! a^{\dag}_{-(k-p), \tau'_{z}}\theta 
\right)
\!+\! 
O \left( \! {\displaystyle \frac{1}{A}} \! \right)  \\
\\[-8pt]
\cong
\theta \overline{\theta}
{\displaystyle \frac{i \hbar^{3}}{m \sqrt A }} \!
\sum_{p, \tau_{z}} \!
p^{2}
a^{\dag}_{p-\frac{k}{2}, \tau_{z}} 
a_{p+\frac{k}{2}, \tau_{z}}
\!-\!
\theta \overline{\theta}
{\displaystyle \frac{i \hbar^{3} k^2}{4 m}}
\rho^{0,0}_{-k}
\!+\! 
{\displaystyle \frac{i \hbar^{3}}{4 m \sqrt A}} \!
\sum_{p \ne 0, p \ne k} 
p(k\!-\!p)
\rho^{0,0}_{-p} \rho^{0,0}_{-(k-p)} \\
\\[-10pt]
+ 
\theta \overline{\theta}
{\displaystyle \frac{i \hbar^{3}}{ m \sqrt A}} 
\sum_{p, \tau_{z} \ne \tau'_{z}} \!
\left( \!
p^{2}
\!-\!
{\displaystyle \frac{k^{2}}{4 }} \!
\right)
a^{\dag}_{p-\frac{k}{2}, \tau_{z}} 
a_{p+\frac{k}{2}, \tau'_{z}}
\!-\!
\theta \overline{\theta}
{\displaystyle \frac{i \hbar^{3}}{ m \sqrt A}} 
\sum_{p} \!
p^{2}
\!+\! 
O \left( \! {\displaystyle \frac{1}{A}} \! \right) ,
\EA
\label{approxrhoPiPi}  
\ea
in the last line of the above Eq.
(\ref{approxrhoPiPi}),
we have used the relations
(\ref{approxa0})
and
the explicit expression for
$\rho^{0,0}_{-k}$
given by the first equation of
(\ref{FcomponentDensityOpT0T1}).

Putting the calculated results of
(\ref{commupiT})
and
(\ref{approxrhoPiPi})
into the
${f}_{k} (\rho)$
(\ref{fkrho})
and discarding the first term in the last line of the Eq.
(\ref{approxrhoPiPi}),
we can obtain an approximate expression for the operator-valued 
function ${{\mit f}_{k}(\rho)}$ up to the order of $\frac{1}{\sqrt A}$
in the following form:
\ba
\BA{c}
{f}_{k} (\rho)
\!=\!
- 
{\displaystyle {\frac{i{\hbar }^{3}{k}^{2}}{4m}}}
\rho^{0,0}_{-k} \ 
\!+\! 
{\displaystyle {\frac{ i{\hbar }^{3}}{4 m \sqrt {A} k}}} \!
\sum_{p \ne 0, p \ne k} 
p (k^{2} \!-\! pk)
\rho^{0,0}_{-p} ~\! \rho^{0,0}_{p-k}
\!-\!
{\displaystyle \frac{i \hbar^{3}}{ m \sqrt A}}
\sum_{p} 
p^{2}
\!+\! 
O \left( \! {\displaystyle \frac{1}{A}} \! \right) ,
\EA
\label{approxfkrho}  
\ea
if the Grassmann variabls
$\theta$ and $\overline{\theta}$
satisfy a condition
$\theta \overline{\theta} \!=\! 1$.
$\!\!$This is just the same form as that obtained by Sunakawa 
\cite{SYN.62}
except that there exists the last constant term
proportional to
$\sum_{p}p^{2}\!$. 
The condition
$\theta \overline{\theta} \!=\! 1$,
however,
is not good for a realistic Fermi system
because the occupation number of the zero-momentum state
$\!\sum_{p,\tau_{z}} \!
a^{\dag}_{p,\tau_{z}} \! a_{p,\tau_{z}}(p \!=\! 0)$
is not equal to the total number $A$ of the N-P system
under consideration.
This is quite a different situation from that
in the Bose system treated by Sunakawa. 
The first idea of regarding the operators
$a_{k=0}$ and $a^{\dag}_{k=0}$
as a $c$-number
$\sqrt N$ (This $N$ is the total number of Bose particles)
at extremely low temperature
was proposed by Bogoliubov
\cite{Bogoliubov.47}. 
For the present moment, 
we can't help using this condition.
Then substituting 
(\ref{approxfkrho}) into 
(\ref{CRPiT0}), 
we can rewrite the RHS of the discrete integral Eq.
(\ref{CRPiT0}) as
\ba
\!\!\!\!\!\!
\BA{c} 
[\Pi^{0,0}_{k} , {T}_{0} (\rho)]
\!=\! 
- {\displaystyle {\frac{i{\hbar }^{3}{k}^{2}}{4m}}} \!
\rho^{0,0}_{-k}
\!+\! 
{\displaystyle {\frac{i{\hbar }^{3}}{4 m \sqrt {\!A} k}}} \!
\sum_{p \ne 0, p \ne k} \!
p({k}^{2} \!\!-\! pk \!+\! {p}^{2})
\rho^{0,0}_{-p} ~\! \rho^{0,0}_{p-k} 
\!+\!
O \! \left( \! {\displaystyle \frac{1}{A \! \sqrt {\!A}}} \! \right) \! .~\!
(k \!\ne\! 0)
\EA
\label{CRPiT02}  
\ea
From
(\ref{CRPiT02}) and the commutation relations
(\ref{CRpirho}) and 
(\ref{exactCRs}),
we get 
\ba
\left.
\BA{ll}
&[\Pi^{0,0}_{k'} , [\Pi^{0,0}_{k} , {T}_{0} (\rho)]]
\!=\! 
-\! 
{\displaystyle {\frac{{\hbar }^{4}{k}^{2}}{4m}}}
{\delta }_{k', -k} 
\!+\! 
{\displaystyle {\frac{{\hbar }^{4}}{4 m \sqrt {A}}}}
({k}^{2}+kk'+{k'}^{2})
\rho^{0,0}_{-k-k'} , \\
\\[-10pt]
&[\Pi^{0,0}_{k''}, [\Pi^{0,0}_{k'} , [\Pi^{0,0}_{k} , {T}_{0}(\rho)]]]
\!=\! 
-\! 
{\displaystyle {\frac{i{\hbar }^{5}}{4 m \sqrt {A}}}}
({k}^{2}+kk'+{k'}^{2})
{\delta }_{k'', -k-k'} , \\
\\[-4pt]
&[\Pi^{0,0}_{k'''} , [\Pi^{0,0}_{k''}, [\Pi^{0,0}_{k'} , [\Pi^{0,0}_{k} , 
{T}_{0} (\rho )]]]]
\!=\!  
0  .
\EA
\right\}
\label{CRPiPiPiPiT0}  
\ea

By a procedure similar to the one in Sec. 4,
we determine the coefficients ${C_{n} (n \!\neq\! 0)}$ and
then get an approximate form of ${T_{0}(\rho)}$
in terms of variables $\rho^{0,0}_{k}$ in the following form:
\ba
\!\!\!\!\!\!
\BA{lll}
{T}_{0} (\rho)
\!\!\!&=&\!\!\!\!\!
{C}_{0} 
\!+\!
{\displaystyle {\frac{{ \hbar }^{2}}{8m}}}
\sum_{p \ne 0} 
{p}^{2} \!
\rho^{0,0}_{p} \! \rho^{0,0}_{-p} \\
\\[-8pt]
\!\!\!&~~-&\!\!\!
{\displaystyle {\frac{{\hbar }^{2}}{24 m \! \sqrt {\!A}}}}
\sum_{p \ne 0, q \ne 0, p+q \ne 0} 
({p}^{2} \!+\! pq \!+\! {q}^{2})
\rho^{0,0}_{p} \rho^{0,0}_{q} \rho^{0,0}_{-p-q}
\!+\! 
O \! \left( \! {\displaystyle \frac{1}{A}} \! \right) \! . 
\EA
\label{T0rho}  
\ea
Using the following identities:
\ba
\left.
\BA{c}
\sum_{p \ne 0, q \ne 0, p+q \ne 0}
{p}^{2}
\rho^{0,0}_{p} \rho^{0,0}_{q}
\rho^{0,0}_{ -p-q} 
\!=\!
\sum_{p \ne 0, q \ne 0, p+q \ne 0} 
(p\!+\!q)^{2} \!
\rho^{0,0}_{p} \rho^{0,0}_{q} \rho^{0,0}_{-p-q}   ,\\
\\
\sum_{ p \ne 0, q \ne 0, p+q \ne 0} 
{p}^{2}
\rho^{0,0}_{p} \rho^{0,0}_{q} \rho^{0,0}_{ -p-q} 
\!=\! 
- 2 
\sum_{ p \ne 0, q \ne 0, p+q \ne 0} 
pq
\rho^{0,0}_{p}  \rho^{0,0}_{q} \rho^{0,0}_{-p-q}  ,
\EA
\right\}
\label{identity}  
\ea
the lowest kinetic term  of $T$,
${T}_{0} (\rho)$
(\ref{T0rho})
is rewritten as
\ba
\BA{c}
{T}_{0} (\rho)
\!=\! 
{C}_{0} 
\!+\!
{\displaystyle {\frac{{\hbar }^{2}}{8m}}}
\sum_{p \ne 0} {p}^{2}
\rho^{0,0}_{p}\rho^{0,0}_{-p} 
\!+\!
{\displaystyle {\frac{{\hbar }^{2}}{8 m \sqrt {A}}}}
\sum_{p \ne 0, q \ne 0, p+q \ne 0} 
pq
\rho^{0,0}_{p} \rho^{0,0}_{q} \rho^{0,0}_{-p-q}  ,
\EA
\label{T0rho2}  
\ea
where
the constant term $C_{0}$ remains undetermined yet.
Then the detailed calculation 
of $C_{0}$ will be given in the next section.


\newpage

\def\thesection{\arabic{section}}
\setcounter{equation}{0}
\renewcommand{\theequation}{\arabic{section}.\arabic{equation}}     
\section{Calculation of the Constant Term}
\hspace{15pt}
In this section, 
we calculate the constant term $C_{0}$, 
the first term in the RHS of
(\ref{T0rho2}).
Substituting
(\ref{T0rho2})  
into
(\ref{exactTPi}),
the constant term $C_{0}$ is computed up to the 
order of $\frac{1}{A}$:\\[-12pt]
\ba
\BA{lll}
{C}_{0} 
=
T
\!\!\!\!&-&\!\!\!\! 
{\displaystyle {\frac{{\hbar }^{2}}{8m}}} \!
\sum_{k} \!
{k}^{2} \!
\rho^{0,0}_{k} \rho^{0,0}_{-k} 
\!-\! 
{\displaystyle {\frac{1}{2m}}} \!
\sum_{k} \!
{k}^{2}
\Pi^{0,0}_{k} \Pi^{0,0}_{-k} 
\!+\!
{\displaystyle {\frac{1}{2 m \! \sqrt {\!A}}}} \!
\sum_{k \ne 0, p \ne 0, k+p \ne 0} \!
kp
\rho^{0,0}_{k+p} \Pi^{0,0}_{k} \Pi^{0,0}_{p} \\
\\[-12pt]
\!\!\!\!&-&\!\!\!\!
{\displaystyle {\frac{{\hbar }^{2}}{8 m \sqrt {A}}}}
\sum_{k \ne 0, p \ne 0, k+p \ne 0}
kp
\rho^{0,0}_{-k-p} \rho^{0,0}_{k} \rho^{0,0}_{p} .
\EA
\label{C0phiexpansion}
\ea\\[-8pt]
Using
$\Pi^{0,0}_{k} \!\cong\!  \pi^{0,0}_{k}$,
$\!\rho^{0,0}_{k}
\!\cong\!
\sum_{\tau_{z}} \!\!
\left( \!
\overline{\theta}a_{-k, \tau_{z}}
\!+\!
a^{\dag}_{k, \tau_{z}}\theta \!
\right)
$
and
$\pi^{0,0}_{k}
\!\cong\!
-{\displaystyle \frac{i\hbar}{2}} \!
\sum_{\tau_{z}} \!\!
\left( \!
\overline{\theta}a_{k, \tau_{z}}
\!-\!
a^{\dag}_{-k, \tau_{z}}\theta \!
\right)
$,
first we can calculate the third term in
(\ref{C0phiexpansion})
similarly to the calculation of
(\ref{approxrhoPiPi}) 
and get a result as \\[-14pt]
\ba
\BA{c}
-
{\displaystyle \frac{1}{2 m}} \!
\sum_{k} 
k^{2}
\Pi^{0,0}_{k} \Pi^{0,0}_{-k}
\!\cong\!
-
{\displaystyle \frac{1}{2 m}} \!
\sum_{k} 
k^{2}
\pi^{0,0}_{k} \pi^{0,0}_{-k}  \\
\\[-12pt] 
\!=\!
{\displaystyle \frac{{\hbar }^{2}}{8 m}} \!
\sum_{k} 
k^{2} \!
\sum_{\tau_{z}} \!
\left( 
\overline{\theta}a_{k, \tau_{z}} 
\!-\!
a^{\dag}_{-k, \tau_{z}}\theta 
\right) \!
\sum_{\tau'_{z}} \!
\left( 
\overline{\theta}a_{-k, \tau'_{z}} 
\!-\!
a^{\dag}_{k, \tau'_{z}}\theta 
\right)    \\
\\[-8pt] 
\!=\!
{\displaystyle \frac{{\hbar }^{2}}{8 m}} \!
\sum_{k} 
k^{2} \!
\left( 
\rho^{0,0}_{-k} 
\!-\!
2 \! \sum_{\tau_{z}} \! a^{\dag}_{-k, \tau_{z}}\theta 
\right) \!
\left( 
\rho^{0,0}_{k}
\!-\!
2 \! \sum_{\tau'_{z}} \! a^{\dag}_{k, \tau'_{z}}\theta 
\right) \\
\\[-10pt]
\!=\! 
{\displaystyle \frac{{\hbar }^{2}}{8 m}} \!
\sum_{k} \! 
k^{2}
\rho^{0,0}_{k}\rho^{0,0}_{-k}
\!-\! 
\theta \overline{\theta} 
\sum_{k} \!
{\displaystyle \frac{{\hbar }^{2} k^{2}}{2 m}} \! 
\sum_{ \tau_{z}, \tau'_{z}} \!
a^{\dag}_{k, \tau_{z}} 
a_{k, \tau'_{z}}
\!+\!
\theta \overline{\theta}
{\displaystyle \frac{{\hbar }^{2}}{2 m}} \!
\sum_{k} \!
k^{2}
\!+\!
O \left( \! {\displaystyle \frac{1}{A}} \! \right)  .
\EA
\label{approxPiPi2}  
\ea
 
As for the case of the forth and fifth terms in
(\ref{C0phiexpansion}),
contrast to the case of the first term, due to the properties
$\theta \theta \!=\! 0$ and $\overline{\theta} \overline{\theta} \!=\! 0$
we simply have 
\ba
\!\!\!\!\!\!\!\!
\BA{ll}
&~~
{\displaystyle {\frac{1}{2 m \! \sqrt {\!A}}}} \!
\sum_{k \ne 0, p \ne 0, k+p \ne 0} \!
kp
\rho^{0,0}_{k+p} \pi^{0,0}_{k} \pi^{0,0}_{p} \\
\\[-8pt]
&\!\cong\! 
-
{\displaystyle {\frac{{\hbar }^{2}}{8 m \! \sqrt {\!A}}}} \!
\sum_{k \ne 0, p \ne 0, k+p \ne 0} \!
kp \!
\sum_{\tau_{z}} \!
\left( \!
\overline{\theta}\!a_{-k-p, \tau_{z}} 
\!\!+\!\!
a^{\dag}_{k+p, \tau_{z}}\!\theta \!
\right) \!
\sum_{\tau'_{z}} \!
\left( \!
\overline{\theta}\!a_{k, \tau'_{z}} 
\!\!-\!\!
a^{\dag}_{-k, \tau'_{z}}\!\theta \!
\right) \!
\sum_{\tau''_{z}} \!
\left( \!
\overline{\theta}\!a_{p, \tau''_{z}} 
\!\!-\!\!
a^{\dag}_{-p, \tau''_{z}}\!\theta \!
\right)  \\
\\[-10pt]
& 
\!=\! 
0  ,
\EA
\label{rhopipi}
\ea\\[-8pt]
and\\[-14pt] 
\ba
\!\!\!\!\!\!\!\!
\BA{ll}
&-
{\displaystyle {\frac{{\hbar }^{2}}{8 m \! \sqrt {\!A}}}} \!
\sum_{k \ne 0, p \ne 0, k+p \ne 0} \!
kp
\rho^{0,0}_{-k-p} \rho^{0,0}_{k} \rho^{0,0}_{p} \\
\\[-8pt]
&\!\cong\! 
-
{\displaystyle {\frac{{\hbar }^{2}}{8 m \! \sqrt {\!A}}}} \!
\sum_{k \ne 0, p \ne 0, k+p \ne 0} \!
kp \!
\sum_{\tau_{z}} \!
\left( \!
\overline{\theta}\!a_{k+p, \tau_{z}} 
\!\!+\!\!
a^{\dag}_{-k-p, \tau_{z}}\!\theta \!
\right) \!
\sum_{\tau'_{z}} \!
\left( \!
\overline{\theta}\!a_{-k, \tau'_{z}} 
\!\!+\!\!
a^{\dag}_{k, \tau'_{z}}\!\theta \!
\right) \!
\sum_{\tau''_{z}} \!
\left( \!
\overline{\theta}\!a_{-p, \tau''_{z}} 
\!\!+\!\!
a^{\dag}_{p, \tau''_{z}}\!\theta \!
\right)  \\
\\[-10pt]
& 
\!=\! 
0  .
\EA
\label{rhorhorho}
\ea 
Substituting Eqs.
(\ref{approxPiPi2})
$\!\sim\!$
(\ref{rhorhorho})
into
(\ref{C0phiexpansion}), 
we have a result\\[-14pt]
\ba
\BA{l}
C_{0}
\!\cong\!
\left( 1 \!-\! \theta\overline{\theta} \right) \!
T
\!+\!
\theta\overline{\theta}  
{\displaystyle {\frac{{\hbar }^{2}}{2m}}} \!
\sum_{k} \! {k}^{2}
\!-\! 
\theta \overline{\theta} 
\sum_{k, \tau_{z} \ne \tau'_{z}} \!
{\displaystyle \frac{{\hbar }^{2} k^{2}}{2 m}} \! 
a^{\dag}_{k, \tau_{z}} \!
a_{k, \tau'_{z}} \\
\\[-12pt]
~~~
\!\cong\!
{\displaystyle {\frac{{\hbar }^{2}}{2 m}}} \!
\sum_{k} \! {k}^{2} .
\EA
\label{resultC0}
\ea\\[-4pt] 
Here we have used the relation
$\theta \overline{\theta} \!\!=\!\! 1$
and neglected
$
\sum_{k, \tau_{z} \ne \tau'_{z}} \!\!
 \frac{{\hbar }^{2} k^{2}}{2 m} \! 
a^{\dag}_{k, \tau_{z}} \!
a_{k, \tau'_{z}}
$
in the first line
which does not exist
in the case of  the {\it isospin}-less Fermion system.
The result
(\ref{resultC0})
is not identical with the Sunakawa's result
\cite{SYN.62}
for a Bose system.
This is because
we have dealt with a Fermi system.
Then we get a result which is considered
as the natural consequence for a Fermi system.
It is surprising to see that the $C_0$
(\ref{resultC0})
coincides with the constant term in the
resultant ground state energy
given by the Tomonaga's method
\cite{Tomo.50}.


\newpage

\def\thesection{\arabic{section}}
\setcounter{equation}{0}
\renewcommand{\theequation}{\arabic{section}.\arabic{equation}}
\section{Discussions and Further Perspectives}
 
Using
(\ref{exactTPi}), (\ref{T0rho2}) and (\ref{resultC0})
and separating the term
$C_{0} \!\cong\! \sum_{k} \! \frac{{\hbar }^{2}{k}^{2}}{2m}$
into two parts
$-\!\sum_{k} \! \frac{{\hbar }^{2}{k}^{2}}{4m}$
and
$\frac{3}{2}\! \sum_{k} \! \frac{{\hbar }^{2}{k}^{2}}{2m}$
and denoting
$\nu^{F}_{T=0}(k)$ and $\nu_{T=0}(k)$
simply as
$\nu^{F}(k)$ and $\nu(k)$,
we can express the original Hamiltonian $H$ in terms of the 
$exact$ canonical variables
$\rho^{0,0}_{k}$ and $\Pi^{0,0}_{k}$
as\\[-20pt]
\ba
\!\!\!\!
\BA{lll}
H 
\!\!\!&=&\!\!\!\!
-{\displaystyle \frac{A \! \left(A \!+\! 2 \right)}{8L}} 
\nu(0) 
\!-\!
\sum_{k} \!
{\displaystyle \frac{{\hbar }^{2}{k}^{2}}{4m}}
\!-\!
{\displaystyle \frac{A}{4L}} \!
\sum_{k \ne 0} 
\nu(k) \\
\\[-10pt]
&&~~~~~~~~~~~~~~~~~~\!\!+
\sum_{k \ne 0} \!
\left\{ \! 
{\displaystyle \frac{{k}^{2}}{2 m}}
\Pi^{0,0}_{k} \Pi^{0,0}_{-k}
\!+\!
\left( \!
{\displaystyle \frac{{\hbar }^{2}{k}^{2}}{8 m}} 
 \!-\!
{\displaystyle \frac{A}{8L}} \!
\nu(k) \!
\right) \!
\rho^{0, 0}_{k} \! 
\rho^{0, 0}_{-k} \!
\right\} \\
\\[-12pt]
\!\!\!\!&-&\!\!\!\!\! 
{\displaystyle \frac{1}{2 m \! \sqrt {\!A}}} \!\!
\sum_{p \ne 0, q \ne 0, p\!+\!q \ne 0} \!
pq
\rho^{0,0}_{p+q} \Pi^{0,0}_{p} \Pi^{0,0}_{q}
\!\!+\!\! 
{\displaystyle \frac{{\hbar }^{2}}{8 m \! \sqrt {\!A}}} \!\!
\sum_{p \ne 0, q \ne 0, p\!+\!q \ne 0} \!
pq
\rho^{0,0}_{p} \! \rho^{0,0}_{q} \! \rho^{0,0}_{-p-q}  
\!+\!\!
{\displaystyle \frac{3}{2}} \!
\sum_{k} \!\!
{\displaystyle \frac{{\hbar }^{2}{k}^{2}}{2m}}  ,
\EA
\label{exactH}
\ea\\[-14pt]
where we have used the relation
$
\nu^{F}(k)
\!\!=\!\!
- \frac{1}{2} \nu(k)
$.
This is Sunakawa's form up to the order of $\frac{1}{\sqrt A}$
\cite{SYN.62},
except the last term
$
\frac{3}{2} \sum_{k} \frac{{\hbar }^{2}{k}^{2}}{2m} 
$
in the RHS of
(\ref{exactH}).
This difference arises due to the fact that
we deal with a Fermi system
but not a Bose system.
At the present moment,
we discard this term.
The sum of the three terms in the first line
and
the two terms of the second line
in
(\ref{exactH}) 
are considered as the lowest order Hamiltonian
$H_0$ \\[-22pt]
\ba
\!\!\!\!
\BA{c}
H_0
\!=\!
- {\displaystyle \frac{A \! \left(A \!+\!2 \right)}{8L}} 
\nu(0) \!
\!+\!
\sum_{ \! k \ne 0} \!
\left\{ \!
- \displaystyle
{{\frac{{\hbar }^{2}{k}^{2}}{4m}}
\!-\!
\frac{A}{4L}
\nu(k)
\!+\!
{\frac{{k}^{2}}{2m}}
\Pi^{0,0}_{k} \Pi^{0,0}_{-k}
\!+\!
\left( \!
\frac{{\hbar }^{2}{k}^{2}}{8 m}
\!-\!
\frac{A}{8L}
\nu(k) \!
\right) \!
\rho^{0,0}_{k} \!\! \rho^{0,0}_{-k}
} \!
\right\} .
\EA
\label{H0}
\ea\\[-16pt]
Now, let us introduce
the Boson annihilation and creation operators
defined as\\[-20pt]
\ba
\left.
\BA{c}
\alpha_{k}
\!\equiv\!
\sqrt{\!\displaystyle{\frac{ m E_{k}}{2{\hbar }^{2}{k}^{2}}}} 
\rho^{0,0}_{-k}
\!+\!
\displaystyle{\frac{ik}{\sqrt{\!2 m E_{k}}}} \Pi^{0,0}_{k},~
(k \ne 0) , \\
\\[-10pt]
\alpha^{\dag}_{k}
\!\equiv\!
\sqrt{\!\displaystyle{\frac{ m E_{k}}{2{\hbar }^{2} {k}^{2}}}} 
\rho^{0,0}_{k}
\!-\!
\displaystyle{\frac{ik}{\sqrt{\!2 m E_{k}}}} \Pi^{0,0}_{-k},~
(k \ne 0) .
\EA
\right\}
\label{Boson_ops}
\ea\\[-12pt]
Using
(\ref{Boson_ops})
and
(\ref{approxpirho}),
the {\it exact} canonical collective variables
$\rho^{0,0}_{k}$ and $\Pi^{0,0}_{k}$
are expressed as\\[-22pt]
\ba
\left.
\BA{cc}
&\rho^{0,0}_{k}
\!=\!
\sqrt{\!{\displaystyle \frac{{\hbar }^{2}{k}^{2}}{2 m E_{k}}}}
{\displaystyle \frac{1}{2}} \!
\left( 
\alpha_{-k}
\!+\!
\alpha^\dag_{k}
\right)
\!=\!
\sum_{\tau_{z}} \! 
\overline{\theta}a_{-k, \tau_{z}}
\!+\!
\sum_{\tau_{z}} \!
a^{\dag}_{k, \tau_{z}}\theta  , \\
\\[-10pt]
&\Pi^{0,0}_{k}
\!=\!
- i {\displaystyle \frac{\sqrt{2 m E_{k}}}{k}}
{\displaystyle \frac{1}{2}} \!
\left( 
\alpha_{k}
\!-\!
\alpha^\dag_{-k}
\right)
\!=\!
- {\displaystyle \frac{i\hbar}{2}} \!
\left(
\sum_{\tau_{z}} \! 
\overline{\theta}a_{k, \tau_{z}}
\!-\!
\sum_{\tau_{z}} \!
a^{\dag}_{-k, \tau_{z}}\theta 
\right) , ~
(k \ne 0) ,
\EA
\right\}
\label{rhoPi}
\ea\\[-14pt]
substituting which into
(\ref{H0}),
then the lowest order Hamiltonian
$H_{\!0}$
(\ref{H0})
is diagonalized as
\\[-20pt]
\ba
\BA{c}
H_0
=
- {\displaystyle \frac{A \left(A + 2 \right)}{8L}} 
\nu(0)
+
\sum_{k \ne 0} E_{k}
\alpha^{\dag}_{k}\alpha_{k} , ~
E_{k}
\equiv
\sqrt{
\left( 
\varepsilon_{k} 
\right)^2
+
{\displaystyle \frac{{\hbar }^{2}{k}^{2}}{m} 
\frac{A}{2L}}
\nu(k)},  ~
\varepsilon_{k}
\equiv
{\displaystyle \frac{{\hbar }^{2}{k}^{2}}{2m}} .
\EA
\label{diagoH0}
\ea\\[-16pt]
From
(\ref{Boson_ops})-(\ref{diagoH0}),
we have a$\!$ Bogoliubov transformation
for$\!$ Boson-like operators
$\sum_{\tau_{z}} \!
\overline{\theta} a_{k, \tau_{z}}\!$
and
$\sum_{\tau_{z}} \!
a^{\dag}_{k, \tau_{z}} \! \theta$
as \\[-22pt]
\ba
\left.
\BA{cc}
&\alpha_{k}
\!=\!
{\displaystyle \frac{ 1 }{2 \! \sqrt{ \varepsilon_{k} E_{k}}}}
\left[
\left( E_{k}\!+\!\varepsilon_{k}\right) \!
\sum_{\tau_{z}} \! \overline{\theta}a_{k, \tau_{z}}
\!+\!
\left( E_{k}\!-\!\varepsilon_{k}\right) \!
\sum_{\tau_{z}} \! a^{\dag}_{-k, \tau_{z}}\theta 
\right] , \\
\\[-12pt]
&\alpha^{\dag}_{-k}
\!=\!
{\displaystyle \frac{ 1 }{2 \! \sqrt{ \varepsilon_{k} E_{k}}}} \!\!
\left[
\left( 
E_{k}
\!-\!
\varepsilon_{k}
\right) \!
\sum_{\tau_{z}} \! \overline{\theta}a_{k, \tau_{z}}
\!+\!
\left( 
E_{k}
\!+\!
\varepsilon_{k}
\right) \!
\sum_{\tau_{z}} \! a^{\dag}_{-k, \tau_{z}}\theta 
\right] ,
\EA
\right\}
\label{Bogolon_ops}
\ea\\[-14pt]
which is the same as the famous Bogoliubov transformation
for the usual Bosons
\cite{Bogoliubov.47,Uhrenbrock.67}.
The diagonalization
(\ref{diagoH0})
has also been
given for the usual Bose system by Sunakawa
\cite{SYN2.62}. 

To convert the Hamiltonian
(\ref{exactH})
into a coordinate representation,
we introduce collective field variables
$\widehat{\rho}(x)$ and $\widehat{\rho}^{'}(x) $
$\!
(\widehat{\rho} (x) \!\!=\!\! n \!+\! \widehat{\rho}^{'}(x) ,~
n \!\!=\!\! \frac{A}{L})
\!$
and $\Pi (x)$ defined
by the Fourier transformation of the $exact$ canonical variables
$\rho^{0,0}_{k}\!$ and $\Pi^{0,0}_{k}\!$ as\\[-22pt]
\ba
\BA{c}
\widehat{\rho}(x) 
\!\equiv\!  
{\displaystyle \frac{\sqrt A}{L}} 
\sum_{k}
\rho_{k} 
e^{ikx}  ,~ 
\widehat{\rho}^{'}(x)
\!\equiv\! 
{\displaystyle \frac{\sqrt A}{L}} 
\sum_{k \ne 0} 
\rho_{k}
e^{ikx} , ~
\Pi (x)
\!\equiv\! 
{\displaystyle \frac{1}{\sqrt A}} 
\sum_{k \ne 0} \Pi_{k}e^{-ikx} ,
\EA
\label{rhoPiFourier}
\ea
which leads to the following coordinate representation 
of the collective field Hamiltonian:\\[-16pt]
\ba
\!\!\!\!\!\!\!\!
\BA{c}
H
\!\!=\!\!  
\sum_{k} \!\! 
{\displaystyle \frac{{\hbar }^{2} \! {k}^{2}}{2m}} 
\!+\!\!\! 
{\displaystyle \int} \!\! dx V \! (x )  \widehat{\rho} (x)
\!\!+\!\!\! 
{\displaystyle \int} \!\! dx \!\!
\left[ \!
{\displaystyle \frac{m}{2}} \partial _{\!x} \Pi (x)
\!\cdot\!\! \widehat{\rho}(x) \!\!\cdot\!
\partial _{\!x} \Pi (x)
\!\!+\!\!
{\displaystyle \frac{{\hbar }^{2}}{8mn}} \!\!
\left( \!\!
1
\!\!-\!\! 
{\displaystyle \frac{\widehat{\rho}^{'}(x) }{n}}
\!\!+\!\!
{\displaystyle \frac{\widehat{\rho}^{'}(x)^2 }{n^2}} \!
\right) \!
\partial _{\!x} \widehat{\rho} (x) \partial _{\!x} \widehat{\rho}(x) \!
\right] \! ,
\EA
\label{Hcorep}
\ea\\[-12pt] 
which is equal to Sunakawa's result
for a Bose system
except a difference of the first term
$\!
\sum_{k} \!
\frac{{\hbar }^{2}{k}^{2}}{2m}
\!$
from the corresponding term
$\!
-\!  \sum_{k} \!
\frac{{\hbar }^{2}{k}^{2}}{4m}
\!$
given in
\cite{SYN3.62}.
In
(\ref{Hcorep}),
we must emphasize that 
the term
$\!
\sum_{k} \!
\frac{{\hbar }^{2}{k}^{2}}{2m}
\!$
is interpreted as a kinetic energy
in the state of perfect degeneracy
if the term
$
\int \! dx 
\frac{m}{2} \partial _{\!x} \Pi (x)
\!\cdot\!\! \widehat{\rho}(x) \!\!\cdot\!
\partial _{\!x} \Pi (x)
$
is regarded as
the "excitation kinetic energy"
in the sense of Tomonaga
\cite{Tomo.50}.
The collective field Hamiltonian
(\ref{Hcorep})
also was given by 
one of the present author's (S.N.)
in his {\it exact} canonically conjugate momenta approach
to an $SU(N)$ quantum system
\cite{Nish.98}.
Contrary to such collective descriptions,
in the beginning
we already referred to
another way to study of collective motions:
Tomonaga first developed a quite different approach to
an elementary excitation in a Fermi system
\cite{Tomo.50}.
A similar attempt was also  given by Luttinger
\cite{Luttinger.63}.
To solve the  Luttinger's model exactly,
Matias and Lieb made a field theoretical approach
based on the fact that density operators $\rho_k$
{\it define} a Bose  field which is {\it ipso facto}
associated with a Fermi-Dirac field.
They obtained the exact and nontrivial spectrum
\cite{MatiasLieb.65}.
Adding to such a historical achievement,
Sunakawa's method for a Fermi system
\cite{SYN.62}
 may be anticipated to work well for the above mentioned problems.
To carry out such a strategy, following Tomonaga
\cite{Tomo.50},
we separate a density operator $\rho^{0,0}_{k}$ into two parts as
$
\rho^{0,0}_{k}
\!\!=\!\!
\rho^{0,0(\!+\!)}_{k}
\!+\!
\rho^{0,0(\!-\!)}_{k}
$
where
$\rho^{0,0(\!+\!)}_{k}$
and
$\rho^{0,0(\!-\!)}_{k}$
are defined as\\[-16pt] 
\ba
\!\!
\BA{c}
\rho^{0,0(\!+\!)}_{k}
\!\!\equiv\!\!
{\displaystyle \frac{1}{\sqrt A}} \!\!
\sum_{p>0, \tau_{z}} \!\!
a^{\dag}_{p+\frac{k}{2}, \tau_{z}} \!
a_{p-\frac{k}{2}, \tau_{z}}, ~
\rho^{0,0(\!-\!)}_{k}
\!\!\equiv\!\!
{\displaystyle \frac{1}{\sqrt A}} \!\!
\sum_{p<0, \tau_{z}} \!\!
a^{\dag}_{p+\frac{k}{2}, \tau_{z}} \!
a_{p-\frac{k}{2}, \tau_{z}} .
\EA
\label{rhoplusminus}
\ea\\[-12pt]
Further according to Tomonaga
\cite{Tomo.50},
we introduce collective momenta
$\pi^{0,0(\!+\!)}_{k}$
and
$\pi^{0,0(\!-\!)}_{k}$
defined as\\[-20pt]
\ba
{\pi_{k}^{0,0(\!\pm\!)}}
\!\equiv\!
{\displaystyle \frac{m} {{k}^{2}}}
\dot{ \left(\rho_{-k}^{0,0(\!\pm\!)}\right) }
\!=\!
{\displaystyle \frac{m} {{k}^{2}}}
{\displaystyle \frac{i} {\hbar}}
[H, \rho^{0,0(\!\pm\!)}_{-k} ]
\!=\!
{\displaystyle \frac{m} {{k}^{2}}}
{\displaystyle \frac{i} {\hbar}}
[T, \rho^{0,0(\!\pm\!)}_{-k} ]  ,
~(k \!\ne\! 0) ~,
\label{piplusminusk} 
\ea\\[-14pt]
and the momentum operator $\pi^{0,0}_{k}$ is expressed as
$
\pi^{0,0}_{k}
\!\!=\!\!
\pi^{0,0(\!+\!)}_{k}
\!+\!
\pi^{0,0(\!-\!)}_{k}
$.
Calculating the commutator (\ref{piplusminusk}),
we obtain the explicit 
expressions for the collective variables
$\pi_{k}^{0,0(\!\pm\!)}$
as \\[-18pt]
\ba
\BA{c}
\!\!\!\!
{\pi_{k}^{0,0(\!+\!)}}
\!\!=\!\!
-{\displaystyle \frac{i \hbar}{\sqrt A k}} \!\!
\sum_{p>0, \tau_{z}} \!
p
a^{\dag}_{p-\frac{k}{2}, \tau_{z}} \!
a_{p+\frac{k}{2}, \tau_{z}} , ~{\pi_{k}^{0,0(\!-\!)}}
\!\!=\!\!
-{\displaystyle \frac{i \hbar}{\sqrt A k}} \!\!
\sum_{p<0, \tau_{z}} \!
p
a^{\dag}_{p-\frac{k}{2}, \tau_{z}} \!
a_{p+\frac{k}{2}, \tau_{z}} .
\EA
\label{piplusminusk2}  
\ea \\[-16pt]
Under the above preliminaries for
introducing the collective variables,
an $exact$ canonical momenta approach
to the one-dimensional neutron-proton system may be developed.
More specifically,
the interesting problem of describing
the elementary excitations occurring in a rod at  $isospin~T \!=\! 0$,
the dipole oscillations of one-dimensional nuclei,
the Goldhaber-Teller model
\cite{GT.48}
or
the Steinwedel-Jensen model
\cite{SJ.50}
may be considered.
See also Refs.
\cite{RS.80}
and
\cite{EG.87}.
By applying the $exact$ canonical momenta approach to these models,
a better description of the elementary energy excitations is expected to be obtained,
because that approach is designed to take
into account essential many-body effects, which were not considered
in previous treatments of these models. In this context, a new field
of exploration of elementary excitations of a one-dimensional Fermi
system, which is intended to be presented elsewhere, may be opened.

Finally, we emphasize the possibility of extending
the present $exact$ canonical momenta approach to the three-dimensional case
which would lead to the $isospin~T \!=\! 0$ quantum hydrodynamics,
by introducing the velocity operator 
$\mbox{\boldmath $v$}_{\mbox{\boldmath $k$}}$
instead of canonical conjugate momentum
$\mbox{\boldmath $\Pi$}_{\mbox{\boldmath $k$}}$
as has been done by Sunakawa
\cite{SYN2.62,SYN3.62}.


\newpage

\noindent
\centerline{\bf Acknowledgements}

\vspace{0.5cm}

$\!\!\!\!\!\!\!\!\!\!$
One of the authors (S.N.) would like to
express his sincere thanks to
Professor Constan\c{c}a Provid\^{e}ncia for
kind and warm hospitality extended to him at
the Centro de F\'\i sica, Universidade de Coimbra.
This work was supported by FCT (Portugal) under the project
CERN/FP/83505/2008.


\newpage

\leftline{\large{\bf Appendix}} 
\appendix

\def\thesection{\Alph{section}}
\setcounter{equation}{0}  
\renewcommand{\theequation}{\Alph{section}. \arabic{equation}}
\section{Commutators Among
$\rho^{T_{1},T_{z1}}_{k_{1}}\!$
and
$\rho^{T_{2},T_{z2}}_{k_{2}}\!$ and
Derivation of (\ref{CRpirho}) and (\ref{CRpiPi})}

~~
Using the first relation of
(\ref{FcomponentDensityOp}),
commutators among
$\rho^{T_{1},T_{z1}}_{k_{1}}\!$
and
$\rho^{T_{2},T_{z2}}_{k_{2}}\!$
are calculated as
\ba
\!\!\!\!\!\!\!\!
\BA{l}
\left[ \rho^{T_{1},T_{z1}}_{k_{1}},
\rho^{T_{2},T_{z2}}_{k_{2}} \right] \\
\\
\!=\!
{\displaystyle \frac{2}{A}} \!
\sum_{p_{1}, \tau_{z1},\tau^{\prime}_{z1}} \!
\sum_{p_{2}, \tau_{z2},\tau^{\prime}_{z2}} \!
\langle
{\displaystyle \frac{1}{2}} \tau_{z1}
{\displaystyle \frac{1}{2}} \tau^{\prime}_{z1}
|T_{1} T_{z1}
\rangle
\langle
{\displaystyle \frac{1}{2}} \tau_{z2}
{\displaystyle \frac{1}{2}} \tau^{\prime}_{z2}
|T_{2} T_{z2}
\rangle
(-1)^{\frac{1}{2} \!+\! \tau^{\prime}_{z1}}
(-1)^{\frac{1}{2} \!+\! \tau^{\prime}_{z2}} \\
\\[-4pt]
\!\times\!
\left( \!
a^{\dag}_{p_{1}+\frac{k_{1}}{2}, \tau_{z1}}
a_{p_{2}-\frac{k_{2}}{2}, -\tau^{\prime}_{z2}}
\delta_{p_{1}-\frac{k_{1}}{2}, p_{2}+\frac{k_{2}}{2}}
\delta_{ -\tau^{\prime}_{z1}, \tau_{z2}}
\!-\!
a^{\dag}_{p_{2}+\frac{k_{2}}{2}, \tau_{z2}}
a_{p_{1}-\frac{k_{1}}{2}, -\tau^{\prime}_{z1}}
\delta_{p_{1}+\frac{k_{1}}{2}, p_{2}-\frac{k_{2}}{2}}
\delta_{\tau_{z1}, -\tau^{\prime}_{z2}} \!
\right)  \\
\\ 
\!=\!
{\displaystyle \frac{2}{A}} \!\!
\sum_{p} \!\!
\sum_{ \tau_{z1},\tau_{z2},\tau^{\prime}_{z2}} \!
\langle
{\displaystyle \frac{1}{2}} \tau_{z1}
{\displaystyle \frac{1}{2}} \!\!-\!\! \tau_{z2}
|T_{\!1} T_{\!z1}
\rangle \!
\langle
{\displaystyle \frac{1}{2}} \tau_{z2}
{\displaystyle \frac{1}{2}} \tau^{\prime}_{z2}
|T_{\!2} T_{\!z2}
\rangle \!
(\!-1\!)^{\frac{1}{2} \!-\! \tau_{z2}}
(\!-1\!)^{\frac{1}{2} \!+\! \tau^{\prime}_{z2}}
a^{\dag}_{p+\frac{k_{1}\!+\!k_{2}}{2}, \tau_{z1}} \!\!
a_{p-\frac{k_{1}\!+\!k_{2}}{2}, -\tau^{\prime}_{z2}} \\
\\[-4pt]
\!-\!
{\displaystyle \frac{2}{A}} \!\!
\sum_{p} \!\!
\sum_{ \tau_{z1},\tau_{z2},\tau^{\prime}_{z2}} \!
\langle
{\displaystyle \frac{1}{2}} \tau_{z1}
{\displaystyle \frac{1}{2}}  \tau'_{z1}
|T_{\!1} T_{\!z1}
\rangle \!
\langle
{\displaystyle \frac{1}{2}} \tau_{z2}
{\displaystyle \frac{1}{2}} \!\!-\!\! \tau_{z1}
|T_{\!2} T_{\!z2}
\rangle \!
(\!-1\!)^{\frac{1}{2} \!+\! \tau^{\prime}_{z1}}
(\!-1\!)^{\frac{1}{2} \!-\! \tau_{z1}}
a^{\dag}_{p+\frac{k_{1}\!+\!k_{2}}{2}, \tau_{z2}} \!\!
a_{p-\frac{k_{1}\!+\!k_{2}}{2}, -\tau^{\prime}_{z1}} \\
\\
\!=\!
{\displaystyle \frac{\sqrt 2}{\sqrt A}} \!
\sum_{T_{3},T_{z3}}
\sqrt{ (2T_{2} \!+\! 1) (2T_{3} \!+\! 1)}
W \!
\left( \!
{\displaystyle \frac{1}{2}} {\displaystyle \frac{1}{2}} T_{2} T_{3};
T_{1} {\displaystyle \frac{1}{2}} \!
\right) \!
\langle
T_{3} T_{z3}
T_{2} \!-\! T_{z2}
|T_{1} \!-\! T_{z1}
\rangle  \\
\\[-4pt]
~~\!\times\!
{\displaystyle \frac{\sqrt 2}{\sqrt A}}
\left\{
\sum_{ p, \tau_{z1}, \tau^{\prime}_{z2}} \! 
\langle
{\displaystyle \frac{1}{2}} \tau_{z1}
{\displaystyle \frac{1}{2}} \tau^{\prime}_{z2}
|T_{3} T_{z3}
\rangle
a^{\dag}_{p+\frac{k_{1}+k_{2}}{2}, \tau_{z1}}
(-1)^{\frac{1}{2} \!+\! \tau^{\prime}_{z2}}
a_{p-\frac{k_{1}+k_{2}}{2}, -\tau^{\prime}_{z2}}
\right. \\
\\[-6pt]
\left.
~~-(-1)^{T_{1} \!+\! T_{2} \!+\! T_{3}} \!
\sum_{ p, \tau_{z2}, \tau^{\prime}_{z1}} \! 
\langle
{\displaystyle \frac{1}{2}} \tau_{z2}
{\displaystyle \frac{1}{2}} \tau^{\prime}_{z1}
|T_{3} T_{z3}
\rangle
a^{\dag}_{p+\frac{k_{1}+k_{2}}{2}, \tau_{z2}}
(-1)^{\frac{1}{2} \!+\! \tau^{\prime}_{z1}}
a_{p-\frac{k_{1}+k_{2}}{2}, -\tau^{\prime}_{z1}}
\right\} \\
\\
\!=\!
{\displaystyle \frac{\sqrt 2}{\sqrt {\!A}}} \!\!
\sum_{T_{3},T_{z3}} \!\!
\left\{ \! (\!-1)^{T_{1} \!+\! T_{2} \!+\! T_{3}} \!\!-\!\! 1 \! \right\} \!\!
\sqrt{ \! (2T_{1} \!\!+\!\! 1) (2T_{2} \!\!+\!\! 1)}
W \!\!
\left( \!
{\displaystyle \frac{1}{2}} {\displaystyle \frac{1}{2}} T_{1} T_{2};\!
T_{3} {\displaystyle \frac{1}{2}} \!
\right) \!\!
\langle
T_{1} T_{z1}
T_{2} T_{z2}
|T_{3} T_{z3}
\rangle \!
\rho^{T_{3},T_{z3}}_{k_{1} \!+\! k_{2}} \! ,
\EA
\label{CRVrhoAppendix}
\ea
which
is just the commutation relation given by
(\ref{CRrho}).

From the expressions for 
$\rho^{0,0}_{k'}$
and
$\pi^{0,0}_{k}$,
the commutators among them can be computed as
\ba
\!\!\!\!\!\!\!\!\!\!\!\!
\left.
\BA{ll}
&[\pi^{0,0}_{k}, \rho^{0,0}_{k'}]
\!=\!
-
{\displaystyle \frac{i\hbar}{A k^{2}}}
\sum_{p, p'}
\sum_{ \tau_{z}} \! 
pk \!
\left( \!
a^{\dag}_{p-\frac{k}{2}, \tau_{z}} a_{p+\frac{k}{2}, \tau_{z}}
\!-\!
a^{\dag}_{p'-\frac{k'}{2}, \tau_{z}} a_{p'+\frac{k'}{2}, \tau_{z}} \!
\right)
\!=\!
-
{\displaystyle \frac{i\hbar }{\sqrt {A}}}
{\displaystyle \frac{k'}{k}}
\rho^{0,0}_{k'-k} , \\
\\
&[\pi^{0,0}_{k} ,\pi^{0,0}_{k'}]
\!=\!
-
{\displaystyle \frac{i\hbar^{2}}{A k^{2} k'^{2}}} \!
\sum_{p, p', \tau_{z}}
pkp'k' \!
\left( \!
a^{\dag}_{p-\frac{k}{2}, \tau_{z}} a_{p+\frac{k}{2}, \tau_{z}}
\!-\!
a^{\dag}_{p'-\frac{k'}{2}} a_{p'+\frac{k'}{2}} \!
\right) \\
\\[-8pt]
&~~~~~~~~~~~~\!=\!
-
{\displaystyle \frac{i\hbar^{2}}{A k^{2} k'^{2}}} \!
\sum_{p, \tau_{z}} \!\!
\left\{ \!
\left( \! p\!-\!{\displaystyle \frac{k}{2}} \! \right) \!\! 
\left( \! p\!-\!{\displaystyle \frac{k'}{2}} \! \right)
\!-\!
\left( \! p\!+\!{\displaystyle \frac{k}{2}} \! \right) \!\!
\left( \! p\!-\!{\displaystyle \frac{k'}{2}} \! \right) \!
\right\} \!
kk'
a^{\dag}_{p-\frac{k+k'}{2}, \tau_{z}} \! a_{p+\frac{k+k'}{2}, \tau_{z}} \\
\\[-8pt]
&~~~~~~~~~~~~\!=\!
-
{\displaystyle \frac{i\hbar^{2}}{\sqrt {A}kk'}}
({k}^{2} \!-\! {k'}^{2})
\pi^{0,0}_{k+k'} .
\EA
\right\}
\label{CRpirhopipi}
\ea

\newpage

\def\thesection{\Alph{section}}
\setcounter{equation}{0}  
\renewcommand{\theequation}{\Alph{section}. \arabic{equation}}
\section{Commutator Among
${\pi^{0,0}_{k}}$
and
${\Pi^{0,0}_{k'}}$
and the Proof of the ${Exact}$ Canonical Commutation Relations
and of the Commutativity of ${\Pi^{0,0}_{k}}$ for Different $k$'s}

~~
The commutator among
${\pi^{0,0}_{k}}$
and
${\Pi^{0,0}_{k'}}$ is calculated tediously but straightforwardly as
\ba
\!\!\!\!\!\!\!\!
\BA{lll}
&&[\pi^{0,0}_{k} , \Pi^{0,0}_{k'}] 
\!=\!
[\pi^{0,0}_{k} , \pi^{0,0}_{k'}]
\!-\! 
{\displaystyle \frac{1}{\sqrt {A} k'}} \!
\sum_{p \ne k'}
p \!
\left\{
\rho^{0,0}_{p-k'} 
[\pi^{0,0}_{k} , \pi^{0,0}_{p}]
\!+\! 
[\pi^{0,0}_{k} , \rho^{0,0}_{p-k'}]
\pi^{0,0}_{p}
\right\} \\
\\[-4pt]
&+&\!\!\!\!
{\displaystyle \frac{1}{\sqrt {\!A} k'}}
{\displaystyle \frac{1}{\sqrt {\!A}}} \!
\sum_{p \ne k', q \ne p} \!
q \!
\left\{ \!
\rho^{0,0}_{p-k'}
\rho^{0,0}_{q-p} 
[\pi^{0,0}_{k} , \rho^{0,0}_{q}]
\!\!+\!\!
\left( \!
\rho^{0,0}_{p-k'}
[\pi^{0,0}_{k} , \pi^{0,0}_{q-p}]
\!+\! 
[\pi^{0,0}_{k} , \rho^{0,0}_{p-k'}]
\rho^{0,0}_{q-p} \!
\right) \!
\pi^{0,0}_{q} \!
\right\}  
\!+\! \cdots \\
\\[-4pt]
\!\!&=&\!\!\!\!
{\displaystyle {\frac{i\hbar }{\sqrt {\!A}k}}}
(k\!+\!k') \!\!
\left\{ \!
\pi^{0,0}_{k+k'}
\!\!-\!\! 
{\displaystyle {\frac{ i\hbar }{\sqrt {\!A}(k\!+\!k')}}} \!\!
\left[ \!
\sum_{p \ne k+k'} \!
p 
\rho^{0,0}_{p-k-k'}
\pi^{0,0}_{p} 
\!\!+\!\!
{\displaystyle {\frac{1}{\sqrt {\!A}}}} \!\!
\sum_{p \ne k+k', q \ne p} \!
q 
\rho^{0,0}_{p-k-k'}
\rho^{0,0}_{q-p}
\pi^{0,0}_{q} 
\right.
\right. \\
\\[-4pt]
&~-\!\!&\!\!
\left.
\left.
{\displaystyle {\frac{1}{\sqrt {\!A}}}}
{\displaystyle {\frac{1}{\sqrt {\!A}}}} \!
\sum_{p \ne k+k', q \ne p, r \ne q} \!
r
\rho^{0,0}_{p-k-k'}
\rho^{0,0}_{q-p}
\rho^{0,0}_{r-q}
\pi^{0,0}_{r}
\right]  \!
\right\}
\!+\!
\cdots  \\
\\[-4pt]
\!\!&=&\!\!\!\!
{\displaystyle {\frac{i\hbar }{\sqrt {\!A}k}}}
(k\!+\!k')
\Pi^{0,0}_{k+k'} .
\EA
\label{CRPipi}
\ea
Thus, we can get 
(\ref{CRpirho})
and
(\ref{CRpiPi}).

The ${exact}$ canonical commutation relations are proved as follows:
First, iterating the discrete integral equation (\ref{exactPi}) and using 
(\ref{CRpirho}), we get
\ba
\BA{lll}
[\Pi^{0,0}_{k}, \rho^{0,0}_{k'}]
\!\!&=&\!\!
[\pi^{0,0}_{k}, \rho^{0,0}_{k'}]
- {\displaystyle {\frac{1}{\sqrt {A}k}}}
\sum_{p \ne k}
\rho^{0,0}_{p-k}
[\pi^{0,0}_{p}, \rho^{0,0}_{k'}] \\
\\[-4pt]
&~+\!\!&\!\!
{\displaystyle {\frac{1}{\sqrt {A}k}} {\frac{1}{\sqrt {A}}}}
\sum_{p \ne k, q \ne p} q
\rho^{0,0}_{p-k}\rho^{0,0}_{q-p} 
[\pi^{0,0}_{q}, \rho^{0,0}_{k'}] \\
\\[-4pt]
&~-\!\!&\!\!
{\displaystyle {\frac{1}{\sqrt {A}k}} {\frac{1}{\sqrt {A}}} 
{\frac{1}{\sqrt {A}}}}
\sum_{p \ne  k, q \ne p, r \ne q} r
\rho^{0,0}_{p-k}\rho^{0,0}_{q-p}\rho^{0,0}_{r-q}
[\pi^{0,0}_{r}, \rho^{0,0}_{k'}] 
+ \cdot \cdot \cdot \\
\\[-4pt]
\!\!&=&\!\! 
- {\displaystyle {\frac{i\hbar k'}{\sqrt {A}k}}}
\rho^{0,0}_{k'-k}
\!+\! 
{\displaystyle {\frac{1}{\sqrt {A}k}}}
\sum_{p \ne k} p
\rho^{0,0}_{p-k}
{\displaystyle {\frac{i\hbar k'}{\sqrt {A}p}}}
\rho^{0,0}_{k'-p} \\
\\[-4pt]
&~-\!\!&\!\!
{\displaystyle {\frac{1}{\sqrt {A}k}} {\frac{1}{\sqrt {A}}}}
\sum_{p \ne k, q \ne p} q
\rho^{0,0}_{p-k}
\rho^{0,0}_{q-p}
{\displaystyle {\frac{i\hbar k'}{\sqrt {A}q}}}
\rho^{0,0}_{k'-q} \\
\\[-4pt]
&~+\!\!&\!\! 
{\displaystyle {\frac{1}{\sqrt {A}k}} {\frac{1}{\sqrt {A}}} 
{\frac{1}{\sqrt {A}}}}
\sum_{p \ne k, q \ne p, r \ne q} r
\rho^{0,0}_{p-k}
\rho^{0,0}_{q-p}\rho^{0,0}_{r-q}
{\displaystyle {\frac{i\hbar k'}{\sqrt {A}r}}}
\rho^{0,0}_{k'-r}  
- \cdot \cdot \cdot  \\
\\[-4pt]
\!\!&=&\!\!
- {\displaystyle {\frac{i\hbar k'}{\sqrt {A}k}}}
\rho^{0,0}_{k'-k} 
\!+\! 
{\displaystyle {\frac{i\hbar k'}{\sqrt {A}k}}}
\rho^{0,0}_{k'-k}
(1 - {\delta }_{kk'}) \\
\\[-4pt]
&~+\!\!&\!\!
{\displaystyle {\frac{i\hbar k'}{\sqrt {A}k}} {\frac{1}{\sqrt {A}}}}
\sum_{p \ne k,\ k' \ne p}
\rho^{0,0}_{p-k}\rho^{0,0}_{k'-p} 
- {\displaystyle {\frac{i\hbar k'}{\sqrt {A}k}} {\frac{1}{\sqrt {A}}}}
\sum_{p \ne k,\ k' \ne p} 
\rho^{0,0}_{p-k}\rho^{0,0}_{k'-p} 
- \cdot \cdot \cdot \\
\\[-4pt]
\!\!&=&\!\!
- {\displaystyle {\frac{i\hbar k'}{\sqrt {A}k}}} 
 \rho^{0,0}_{k'-k}
 \cdot {\delta }_{ kk'}
 \!=\!
 - i\hbar {\delta }_{kk'} , ~
(\rho^{0,0}_{0} \!=\! \sqrt {A} )  .
\EA
\label{CRPirho}
\ea
In each summation of the above,
we separate the part with $\!p \!=\! k'\!$ and that with $\!p \!\ne\! k'\!$ 
and use the relation $\rho^{0,0}_{0} \!=\!\! \sqrt {{\!A}}$.
All terms which involve 
higher order powers of ${\rho^{0,0}_{k}}$ cancel out except the 
${c}$-number term in the last line of the RHS of Eq.
(\ref{CRPirho})
which arises due to the exclusion of the $p \!=\! k'$ term
if  $k \!=\! k'$ in the summation with respect to $p$. 
Thus we could reach the ${exact}$ canonical commutation relation. 
In order to assert that
the operators ${\Pi^{0,0}_{k}}$'s are ${exact}$
canonically conjugate to ${\rho^{0,0}_{k}}$'s,
we must give the proof of the commutativity of 
the ${\Pi^{0,0}_{k}}$'s for different ${k}$'s.
The commutation relation among 
the ${\Pi^{0,0}_{k}}$'s
can be calculated tediously but straightforwardly as
\ba
\BA{lll}
[\Pi^{0,0}_{k} , \Pi^{0,0}_{k'}] 
\!\!&=&\!\!
-{\displaystyle {\frac{i\hbar }{\sqrt {A}kk'}}}
({k}^{2}- {k'}^{2})
\pi^{0,0}_{k+k'}
\!+\! 
{\displaystyle {\frac{ i\hbar }{\sqrt {A}kk'}}}
({k}^{2}- {k'}^{2})
\Pi^{0,0}_{k+k'} \\
\\[-4pt]
&~-\!\!&\!\!
{\displaystyle \frac{1}{\sqrt {A}k}}
\sum_{p \ne k} p
\rho^{0,0}_{p-k} 
[\Pi^{0,0}_{p} , \pi^{0,0}_{k'}]
\!+\! 
{\displaystyle \frac{1}{\sqrt {A} k'}}
\sum_{q \ne k'} q
\rho^{0,0}_{q-k'} 
[\Pi^{0,0}_{q} , \pi^{0,0}_{k'}] \\
\\[-4pt]
&~+\!\!&\!\!
{\displaystyle {\frac{1}{\sqrt {A}k}} {\frac{1}{\sqrt {A}k'}}}
\sum_{p \ne k} 
\sum_{q \ne k'} pq
\rho^{0,0}_{p-k} \rho^{0,0}_{q-k'}
[\Pi^{0,0}_{p} , \Pi^{0,0}_{q}]  \\
\\[-4pt]
\!\!&=&\!\!
-{\displaystyle {\frac{i\hbar }{\sqrt {A}kk'}}}
({k}^{2}- {k'}^{2})
\pi^{0,0}_{k+k'}
\!+\! 
{\displaystyle {\frac{ i\hbar }{\sqrt {A}kk'}}}
({k}^{2}- {k'}^{2})
\Pi^{0,0}_{k+k'} \\
\\[-4pt]
&~+\!\!&\!\!
{\displaystyle {\frac{ i\hbar }{\sqrt {A}k\sqrt {A}k'}}}
\sum_{p \ne k} p(p+k')
\rho^{0,0}_{p-k} \Pi^{0,0}_{p+k'} \\
\\[-4pt]
&~-\!\!&\!\!
{\displaystyle {\frac{ i\hbar }{\sqrt {A}k\sqrt {A}k'}}}
\sum_{q \ne k'}q(q+k)
\rho^{0,0}_{q-k'} \Pi^{0,0}_{q+k} \\
\\[-4pt]
&~+\!\!&\!\!
{\displaystyle {\frac{1}{\sqrt {A} k\sqrt {A}k'}}}
\sum_{p \ne k} \sum_{q \ne k'} 
pq
\rho^{0,0}_{p-k}\rho^{0,0}_{q-k'}
[\Pi^{0,0}_{p} , \Pi^{0,0}_{q}] \\
\\[-4pt]
&=&
F(k;k')
\!+\! 
{\displaystyle {\frac{1}{\sqrt {A} k\sqrt {A}k'}}}
\sum_{p \ne k} 
\sum_{q \ne k'} pq
\rho^{0,0}_{p-k} \rho^{0,0}_{q-k'} F(q;p)
\!+\!
\cdot \cdot \cdot \ .
\EA
\label{CRPiPi}
\ea
In the above we have introduced the operator-valued function ${F(k;k')}$ 
defined below.
The definition of the function ${F(k;k')}$
shows that it self-evidently vanishes exactly
owing to 
the definition of the ${exact}$ canonically conjugate momenta
${\Pi^{0,0}_{k}}$, i.e.,
(\ref{exactPi}),
\ba
\BA{c}
F(k;k')
\!\equiv\!
-{\displaystyle {\frac{ i\hbar }{\sqrt {A}kk'}}}
\left( {k}^{2} \!-\! {k'}^{2} \right) \!
\left\{
\pi^{0,0}_{k+k'}
\!-\!
{\displaystyle {\frac{ i\hbar }{\sqrt {A}(k+k')}}}
\sum_{ p~all}^{} p
\rho^{0,0}_{p-(k+k')} \Pi^{0,0}_{p} 
\right\} \\
\\[-4pt]
\!=\! 
-{\displaystyle {\frac{ i\hbar }{\sqrt {A}kk'}}}
\left( {k}^{2} - {k'}^{2} \right) \!
\left\{
\pi^{0,0}_{k+k'}
\!-\!
{\displaystyle {\frac{ i\hbar }{\sqrt {A}(k+k')}}}
\sum_{p \ne k+k'} 
p
\rho^{0,0}_{p-(k+k')} \Pi^{0,0}_{p} \!-\! \Pi^{0,0}_{k+k'}
\right\} \\
\\[-4pt]
\!=\! 
-{\displaystyle {\frac{i\hbar }{\sqrt {A}kk'}}}
\left( {k}^{2} - {k'}^{2} \right) \!
\left\{ \Pi^{0,0}_{k+k'} \!-\! \Pi^{0,0}_{k+k'} \right\}
=
0 .
\label{operatorvaluedF}
\EA
\ea
Thus, we can give the proof of the commutativity of 
the ${\Pi^{0,0}_{k}}$'s for different ${k}$'s.



\newpage


\begin{thebibliography}{999}
\bibitem{SJ.80a}
B. Sakita, 
{\it Phys. Rev.} {\bf D21} (1980) 1067.
\bibitem{SJ.80b}
A. Jevicki and B. Sakita, 
{\it Nucl. Phys.} {\bf B165} (1980) 511.
\bibitem{SJ.80c}
A. Jevicki and B. Sakita,
{\it Nucl. Phys.} {\bf B185} (1981) 89.
\bibitem{SJ.80d}
B. Sakita, {\it Quantum Theory of Many-Variable Systems and Fields},
World Scientific Lecture Notes in Physics 1 (1985).
\bibitem{INTUW.93}
The seminar and workshop on  Large-Amplitude Collective Motion  
held at the National Institute for Nuclear Theory, 
University of Washington, Seattle, 1993.
\bibitem{Nishi.94}
S. Nishiyama,
{\it Nucl. Phys.} {\bf A576} (1994) 317.
\bibitem{Tomo.55a}
S. Tomonaga, 
{\it Prog. Theor. Phys.} {\bf 13} (1955) 467.
 \bibitem{Tomo.55b}
S. Tomonaga, 
{\it Prog. Theor. Phys.} {\bf 13} (1955) 482.
\bibitem{SYN.62}
S. Sunakawa, Y. Yoko-o and H. Nakatani, 
{\it Prog. Theor. Phys.} {\bf 27} (1962) 589.
\bibitem{NishProvi.14}
S. Nishiyama and J. da Provid\^{e}ncia, 
{\it Nucl. Phys.} {\bf A 923} (2014) 51.
\bibitem{Nishi.77}
S. Nishiyama,
{\it Prog. Theor. Phys.} {\bf 58} (1977) 1316.
\bibitem{BM.74}
A. Bohr and B. Mottelson, {\it Nuclear Structure}, Volume II,
W. A. Benjamin, 1974.
\bibitem{UiBi.70a}
H. Ui, {\it Prog. Theor. Phys.} {\bf 44} (1970) 153.
\bibitem{UiBi.70b}
L. Weaver and L.C. Biedenharn, 
{\it Nucl. Phys.} {\bf A185} (1972) 1.
\bibitem{UiBi.70c}
P. Gulshani and D.J. Rowe, 
{\it Can. J. Phys.} {\bf 54} (1976) 970.
\bibitem{NishProviarXive.14}
S. Nishiyama and J. da Provid\^{e}ncia, 
{\it Nucl. Phys.} {\bf A 935} (2015) 1.
\bibitem{BIPZ.78a}
E. Br\`ezin, C. Itzykson, G. Parisi and J.-B. Zuber, {\it Comm. Math. Phys.} 
{\bf 59} (1978) 35.
\bibitem{BIPZ.78b}
C. Itzykson and J.-B. Zuber, 
{\it J. Math. Phys.} {\bf 21} (1980) 411.
\bibitem{Nish.98}
S. Nishiyama,
{\it Int. J. Mod. Phys.}  {\bf A13} (1998) 5535.
\bibitem{Tomo.50}
S. Tomonaga, 
{\it Prog. Theor. Phys.} {\bf 5} (1950) 544.
\bibitem{Emery.79}
V.J. Emery,
Theory of the One-Dimensional Electron Gas
in
{\it Highly Conducting One-Dimensional Solids},
Editors, J.T. Devreese, R.P. Evrard and V.E. van Dore,
Physics of Solids and Liquids,
Springer US, 1979, pp 247-303. 
\bibitem{Luttinger.63}
J.M. Luttinger,
{\it J. Math. Phys.} {\bf 4} (1963) 1154.
\bibitem{Solyon.79}
J. S\'{o}lyon,
{\it Adv. Phys.} {\bf 28} (1979) 201.
\bibitem{Mahan.00}
Gerald D. Mahan,
{\it Many-Particle Physics},
(Physics of solids and liquids),
Kluwer Academic/Plenum Publishers, New York, Third edition, 2000, pp 256-276.
\bibitem{Berezin.66}
F.A. Berezin,
{\it The Method of Second Quantization},
Academic Press, New York and London, 1966.
\bibitem{Casalbuoni.76a}
R. Casalbuoni,
{\it Nuovo Cimento} {\bf 33A} (1976) 115.
\bibitem{Casalbuoni.76b}
R. Casalbuoni,
{\it Nuovo Cimento} {\bf 33A} (1976) 389.
\bibitem{Lipkin.65a}
H.J. Lipkin,
$\!${\it Lie Groups for $\!$Pedestrians} (North-Holland Publ. Co., Amsterdam, 1965).
\bibitem{Lipkin.65b}
S. Goshen and H.J. Lipkin,
in {\it Spectroscopic and group theoretical methods in physics,
Racah Memorial Volume},
Edited by: F. Bloch, S.G. Cohen, A. De-Shalit, S. Sambursky and I. Talmi,
North-Holland Publ. Co., Amsterdam, 1968, pp. 245 - 273.
\bibitem{RoweWood.10}
D.J. Rowe and J.L. Wood,
{\it Fundamentals of Nuclear Models, Foundational Models},
World Scientific Publishing Co. Pte. Ltd., p. 250, 2010.
\bibitem{Bogoliubov.47}
N.N.  Bogoliubov,
{\it J. Phys.} {\bf 11} (1947) 23.
\bibitem{Uhrenbrock.67}
D.A. Uhrenbrock,
{\it Commun. Math. Phys.} {\bf 4} (1967) 64. 
\bibitem{SYN2.62}
S. Sunakawa, Y. Yoko-o and H. Nakatani, 
{\it Prog. Theor. Phys.} {\bf 27} (1962) 600.
\bibitem{SYN3.62}
S. Sunakawa, Y. Yoko-o and H. Nakatani, 
{\it Prog. Theor. Phys.} {\bf 28} (1962) 127. 
\bibitem{MatiasLieb.65}
Daniel C. Matias and Elliott H. Lieb,
{\it J. Math. Phys.} {\bf 6} (1965) 304.
\bibitem{GT.48}
M. Goldhaber and E. Teller,
{\it Phys. Rev.} {\bf 74} (1948) 1046.
\bibitem{SJ.50}
H. Steinwedel and J.H.D. Jesen,
{\it Z. Naturforschung} {\bf 5A} (1950) 413. 
\bibitem{RS.80}
P. Ring and P. Schuck,
{\it The nuclear many body problem},
Texts and monographs in Physics
(Springer-Verlag, Berlin, Heidelberg and New York 1980), p. 558.
\bibitem{EG.87}
J.M. Eisenberg and W. Greiner, 
{\it Nuclear Models}, North-Holland Physics Publishing,
Elsevier Science Publisher Company,  Inc. 1987, p. 585.


\end{thebibliography}
\end{document}